\documentclass[lettersize,journal]{IEEEtran}
\ifCLASSOPTIONcompsoc
  \usepackage[nocompress]{cite}
\else
  \usepackage{cite}
\fi
\ifCLASSINFOpdf
  \usepackage{graphics}
  \usepackage[pdftex]{graphicx}
  \usepackage{amsfonts}
  \usepackage[utf8]{inputenc}
  \usepackage{amsmath}
  \usepackage[ruled,vlined]{algorithm2e}
  \usepackage{blindtext}
  \usepackage{hyperref}
  \usepackage[english]{babel}
  \usepackage[T1]{fontenc}
  \usepackage[utf8]{inputenc}

  \usepackage{booktabs}
  \usepackage{multirow}
  \usepackage{booktabs}
  \usepackage{listings}
  \usepackage{textcomp}
  \usepackage[english]{babel}

\else
\fi
\newcommand{\squishlist}{
  \begin{list}{$\bullet$}
    { \setlength{\itemsep}{0pt}      \setlength{\parsep}{0pt}
      \setlength{\topsep}{0pt}       \setlength{\partopsep}{0pt}
      \setlength{\leftmargin}{1em} \setlength{\labelwidth}{1em}
      \setlength{\labelsep}{0.5em} } }
\newcommand{\squishlisttwo}{
  \begin{list}{$\bullet$}
    { \setlength{\itemsep}{0pt}    \setlength{\parsep}{0pt}
      \setlength{\topsep}{0pt}     \setlength{\partopsep}{0pt}
      \setlength{\leftmargin}{2em} \setlength{\labelwidth}{1.5em}
      \setlength{\labelsep}{0.5em} } }
\newcommand{\squishlistend}{
    \end{list}  }

\begin{document}
%
\title{Pluvio: 
Incorporating Broader Human Knowledge for Assembly Clone Search on Out-of-domain Architectures and Libraries through Large-Scale Language Model 
}

\author{\IEEEauthorblockN{Zhiwei~Fu\IEEEauthorrefmark{1}, Xinyu~Hu\IEEEauthorrefmark{1}, Steven~H.~H.~Ding\IEEEauthorrefmark{1},  Furkan~Alaca\IEEEauthorrefmark{2},
Benjamin~C.~M.~Fung\IEEEauthorrefmark{1},
and~Philippe~Charland\IEEEauthorrefmark{3}} \\
\IEEEauthorblockA{\IEEEauthorrefmark{1}School of Information Studies, McGill University, Montreal, Quebec, Canada.\\
}\IEEEauthorblockA{\IEEEauthorrefmark{2}School of Computing, Queen's University, Kingston, Ontario, Canada. \\
}\IEEEauthorblockA{\IEEEauthorrefmark{3}Mission Critical Cyber Security Section, Defence R\&D Canada - Valcartier, Quebec, QC, Canada. \\ Emails: \{zhiwei.fu, xinyu.hu3\}@mail.mcgill.ca, \{steven.h.ding, ben.fung\}@mcgill.ca, furkan.alaca@queensu.ca,  philippe.charland@drdc-rddc.gc.ca
}}

\IEEEtitleabstractindextext{%
\begin{abstract}
Code reuse is crucial in software development for a faster and more efficient development lifecycle. Assembly clone search has been effective in identifying vulnerable code resulting from reuse in released executables. Recent studies on assembly clone search demonstrate a trend towards using machine learning-based methods to match assembly code variants. However, these methods are limited to what they learn from a small number of toolchain variants used in training, rendering them 
unsuitable for matching code from unseen microarchitectures and compilation toolchain variants.

This paper presents the first study on the problem of assembly clone search that generalizes to unseen microarchitectures and libraries. We propose incorporating human common knowledge into current learning-based approaches for assembly clone search through large language model (LLM). Transfer learning can aid in addressing the limitations of the existing approaches. We further address the sequence limit issue by proposing a reinforcement learning agent to remove unnecessary and redundant tokens. Coupled with a new Variational Information Bottleneck learning strategy, the proposed system minimizes the reliance on potential indicators of microarchitectures and optimization settings. We simulate the unseen microarchitecture clone search scenarios and the experimental results show the effectiveness of the proposed approach against the state-of-the-art solution.

\end{abstract}

}

\maketitle

\IEEEdisplaynontitleabstractindextext

%
\IEEEpeerreviewmaketitle

\section{Introduction}

Code reuse is crucial to software development~\cite{sojer_henkel_2010, krueger_1992}. This involves using parts of software that encapsulate functionality or lines of code from pre-existing systems~\cite{krueger_1992}. High-quality code reuse policies can result in faster and more efficient development, as well as more maintainable software~\cite{sojer_henkel_2010}. In reality, however, many code reuse practices lack proper quality management, resulting in problems such as vulnerability propagation and intellectual property infringement. The complexity of the software supply chain and the widespread use of open-source third-party libraries exacerbate this issue. It is common to find multiple instances of the same vulnerable code across firmware releases\footnote{See: \href{https://binarly.io/posts/The_Firmware_Supply_Chain_Security_is_broken_Can_we_fix_it/index.html}{The Firmware Supply-Chain Security is broken: Can we fix it?}}.

\begin{figure}
  \centering
  \includegraphics[width=\columnwidth]{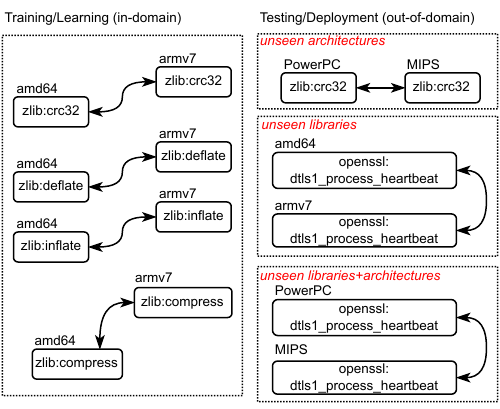}
  \caption{When deploying assembly clone search engines, out-of-domain issues arise. This occurs when certain microarchitectures and libraries are not accessible during the training process, which is limited by access to compilers and the difficulty of encompassing all potential microarchitectures. Our objective is to acquire knowledge from a specific set of microarchitectures and libraries, and then apply it to previously unseen ones.}
  \label{fig:intro}
\end{figure}

Assembly clone search is a critical shift-right defence mechanism that aims to identify vulnerable code resulting from code reuse in released executables. 
It has already proven effective for various security-related tasks, including identifying software plagiarism~\cite{Luo_2017}, detecting injected malware~\cite{Bruschi_2006}, searching for known vulnerabilities in software or firmware images of IoT devices~\cite{Brumley_2008, Ding_2016}, and identifying performance issues, such as increased resource consumption, in patches~\cite{Hu_2016}. A practical clone search system requires a scalable and robust matching approach that can compare diverse assembly code variants corresponding to the same source code, generated by different compilers and with microarchitecture-specific optimizations~\cite{Irfan_2019, Pei_2020, Ding_2016}.

Recent studies on assembly code clone search indicate a trend toward using machine learning-based and large language model (LLM) methods to match assembly code variants produced by out-of-domain toolchains. This trend is due to the inefficiency and limited coverage that result from relying solely on human experts to match assembly code variants produced by different toolchains~\cite{Irfan_2019}. Deep learning models, such as \textsc{Gemini}~\cite{Xu_2017}, \textsc{Asm2Vec}~\cite{Steven_2019}, and the \textsc{Order Matters} model~\cite{Zeping_2020}, are designed to create assembly program embeddings by leveraging assembly functions and Control-Flow Graphs (CFGs). Natural Language Processing (NLP) strategies are employed in \textsc{SAFE}~\cite{Massarelli_2019} and \textsc{PalmTree}~\cite{Li_2021} to improve efficiency and generalizability in cross-platform matching. Pretrained on source code, \textsc{CodeBERT}~\cite{feng_2020} achieves remarkable results on source-level code understanding tasks. General-purpose LLMs, such as \textsc{GPT-5.5}~\cite{openai_gpt55}, show strong reasoning and generation capabilities across a wide range of natural language and programming tasks.

Machine learning models for assembly clone search are generally limited by the small number of toolchain variants that they are trained on. For instance, the \textsc{Order Matters} model~\cite{Zeping_2020} can accurately match x86 and ARM assembly code after training on known pairs of assembly code variants. However, a wider range of toolchain variants exists in the wild, including for microarchitectures such as ARMv4, MIPS series, and Motorola MC68 series. Some hardware vendors, for instance, develop firmware using the same code base for various devices such as webcams, VoIP phones, and wireless routers containing different microarchitectures~\cite{Pewny_2015}. Many of these low-resource or proprietary processor toolchains limit the capacity to generate training data, rendering existing machine learning models inapplicable to unseen microarchitectures and their corresponding compilation toolchains. Moreover, systems such as \textsc{PalmTree} and \textsc{Asm2Vec} are trained and tested on different function pairs from the same set of libraries, leading to performance degradation when making inferences on unseen libraries. We term this practical challenge in assembly clone search \textbf{out-of-domain microarchitectures and libraries}, as illustrated in Figure~\ref{fig:intro}.
This challenge makes assembly code search a difficult problem.

The limitations of current ML-based approaches in generalizing to out-of-domain microarchitectures and libraries stems from their sole reliance on task-specific data, without incorporating the broader knowledge of human experts in assembly code. Consider a reverse engineer who can begin understanding an assembly code fragment of a new microarchitecture without consulting the manual. This individual has a general understanding of how assembly code operates and can identify instructions that are similar to those they already know. Incorporating such expert knowledge can help address the limitations of existing learning-based approaches. A reverse engineer does not learn simply by reading assembly code, but rather through years of education and practical experience. In machine learning, this is akin to the transfer learning paradigm for domain adaptation. Natural Language Processing (NLP) techniques, including ChatGPT models, have achieved tremendous accomplishments in transfer learning. These models excel at understanding natural language text in general and in solving domain-specific tasks, such as programming, through downstream training.

This study simulates real-world scenarios of out-of-domain assembly clone search and explores the potential of using large-scale language models pretrained on general text data. Unlike previous models that solely rely on assembly code, such as the \textsc{Order Matters} model, our study employs models pretrained on natural language text data and treats assembly code as text data. By training on a small number of x86 and ARM assembly code pairs, we find some pretrained models achieve high accuracy (>85\%) in matching unknown microarchitectures, such as MIPS and PowerPC. We further improve performance by addressing two limitations of the pretrained model: \textbf{L1 - sequence length limit} and \textbf{L2 - nuisance information}. 

To tackle L1, we propose the \textit{Removal} module, a reinforcement learning agent that aims to eliminate noise tokens from assembly code. While a typical pretrained natural language model is limited by a sequence length cap of 800 words, an assembly function can be much longer than regular text paragraphs. Function inlining, compiler-specific injected code, and code duplication induced by compiler optimizations expose substantial opportunities for removing redundancies, which, as a result, would enable the model to increase the input sequence length while retaining the crucial elements of the code sequence as the input to the pretrained natural language model. For L2, we propose modifying the Variational Information Bottleneck (VIB) \cite{Tishby_2000, VIB_2017} learning method to compress the information learned about the code's microarchitecture and optimization settings. Since the model should be capable of matching the code of unseen microarchitectures, we treat the assembly code's microarchitectures and optimizations as nuisance information. The goal of the learning is to ``forget'' this information and minimize the input information regarding those factors. 
In this paper, we make the following contributions:
\squishlist
    \item We propose to integrate general human knowledge into the learning process for assembly code similarity. We adopt pretrained natural language models on text data following the transfer learning paradigm, where assembly code is treated as plain text data.
    \item To overcome the sequence length limitation of the pretrained natural language model (L1), we propose using a reinforcement learning agent that is capable of eliminating noisy and redundant tokens via a policy gradient algorithm.
    \item To enhance the model's resilience to out-of-domain microarchitectures and libraries (L2), we introduce a novel approach: Conditional Variational Information Bottleneck (CVIB) learning, where conditions are used only during the training step. This method ensures that the model minimally relies on input information that could serve as indicators of microarchitectures and optimization settings.
    \item We simulate the real-life out-of-domain scenarios and benchmark state-of-the-art methods for clone search. Our result demonstrates the effectiveness of transfer learning, the proposed \textit{Removal} module, and the new CVIB (conditions in training only) approach\footnote{source code and data are publicly available at https://github.com/L1NNA/Pluvio}.
    
\squishlistend

The rest of this paper is structured as follows. Section~\ref{sec:method} elaborates on the general design of \textit{Pluvio}. Section~\ref{sec:exp} presents our experimental setup and results. Section~\ref{sec:related} provides a more thorough review of the relevant state-of-the-art methods for assembly code similarity detection. Finally, Section~\ref{sec:conclude} concludes this paper.

\section{Research Methodology}
\label{sec:method}
\begin{figure*}[h]
  \centering
  \includegraphics[width=0.8\linewidth]{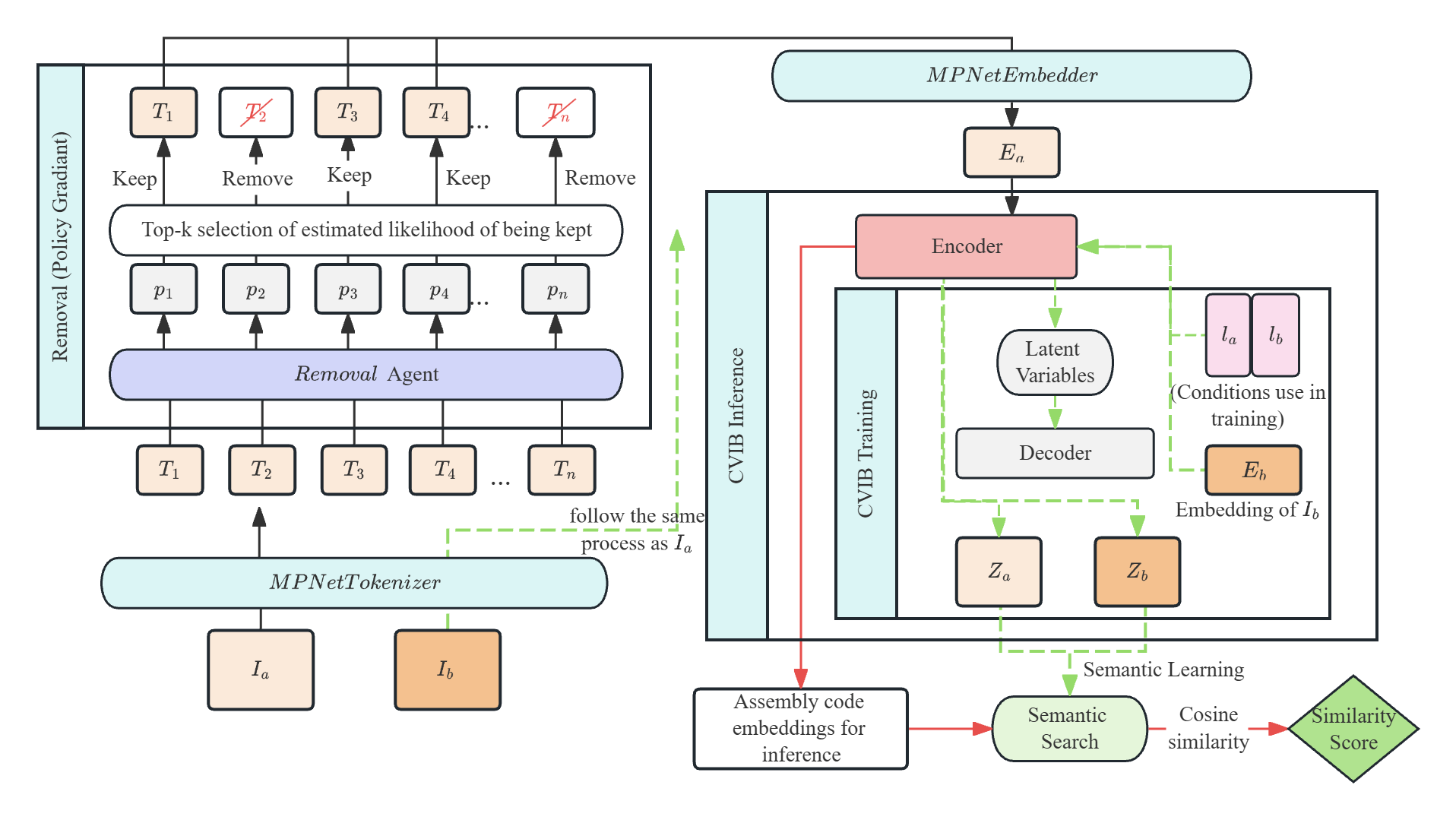}
  \caption{The Structural Overview of \textit{Pluvio}. A pair of assembly instruction sequences $I_a$ and $I_b$ are first tokenized by the MPNetTokenizer. The \textit{Removal} agent estimates a retention probability $p_i$ for each token $T_i$ and performs a top-$k$ selection to retain informative tokens, while removing redundant or noisy tokens. The retained sequences are then encoded by the MPNetEmbedder into function-level embeddings $E_a$ and $E_b$, which are passed to the CVIB Encoder to obtain latent representations. The resulting representations $Z_a$ and $Z_b$ are further optimized through semantic learning. During inference, the CVIB Encoder directly produces the two assembly code embeddings used by semantic search to calculate the similarity score. Green dashed paths denote training-only operations, red paths denote inference-only operations, and solid black paths are shared by both training and inference.}
  \label{fig:pluvio}
\end{figure*}
Our proposed architecture, named \textit{Pluvio}, has three main components. The first component is a large pretrained natural language model, MPNet~\cite{mpnet_2020}, in which there is a tokenizer denoted by \textit{MPNetTokenizer} and an encoder denoted by \textit{MPNetEmbedder}. The language model incorporates enormous human knowledge, pretrained by billions of diverse natural language data samples and then uptrained by pairwise assembly code pairs, to capture the assembly code's semantics. The second component is a reinforcement learning agent model, denoted by \textit{Removal}, to remove noise and reduce the impact from inlined functions and compiler-injected code. The third component is a novel technique that we introduce, designed to remove the nuisance information caused by different microarchitectures, libraries and optimizations, and to achieve better domain adaptation, called Conditional Variational Information Bottleneck (\textit{CVIB}), with conditions in training only. Its encoder and decoder are denoted by \textit{CVIBEncoder} and \textit{CVIBDecoder}.

We train \textit{Pluvio} on a set of assembly function pairs and utilize three different loss functions: cosine similarity loss~\cite{sentence_bert_2019}, denoted by $L_{cos\_sim}$, reinforce learning loss, denoted by $L_{RL}$, and information loss, denoted by $L_{CVIB}$. Finally, we apply $\beta_1$ and $\beta_2$, known as Lagrange parameters, to control the learning rate of $L_{RL}$ and $L_{CVIB}$; thus, the overall loss function $L_{pluvio}$ is defined as 
\begin{equation}
    L_{pluvio} = L_{cos\_sim} + \beta_1 \cdot L_{RL} + \beta_2 \cdot L_{CVIB}.
\end{equation}
Overall, these three parts enable \textit{Pluvio} to capture the maximal semantics and generate the most informative representation of the assembly code.

The following sections thoroughly explain each of these components and Figure \ref{fig:pluvio} visually depicts the \textit{Pluvio} architecture. The input for training \textit{Pluvio} is a pair of assembly instruction sequences denoted by $I_a$ and $I_b$, with the ground truth $y_{actual}$ that is either 1 (similar) or 0 (dissimilar). The output \textit{sscore} is the similarity score of the pair $I_a$ and $I_b$. The goal is to learn a function $f$ that is able to encode $I_a$ and $I_b$ into a pair of fixed-length encodings $Z_a$ and $Z_b$. Then, the function $f: (I_a, I_b) \in \mathbb{D} \rightarrow (Z_a, Z_b) \rightarrow sscore \in \mathbb{R}$ where $sscore \in [0, 1]$ can map two assembly instructions to their similarity score \textit{sscore}. The set $\mathbb{D}$ represents the pairwise out-of-domain testing data from different microarchitectures, libraries, or both (as illustrated in the right half of Figure~\ref{fig:intro}).

\subsection{Pretrained Natural Language Model}
In the NLP domain, the semantic similarity of a pair of texts in natural languages can be determined by calculating the similarity of their embeddings~\cite{Saurabh_2021}. We capture tokens from assembly instructions, rather than from their CFGs and their adjacency matrices, to reduce the effect of noisy information from various microarchitectures, libraries, and compiler optimizations~\cite{Li_2021}. Similar to natural languages, instructions can be treated as a sentence and tokens as words. Thus, we feed our pretrained natural language model a long string (an instruction sequence of one assembly function) as input and then either tokenize or encode it for the \textit{sscore} calculation.

As a breakthrough in language modelling introduced in 2017, a Transformer~\cite{Transformer_2017} consists of a point-wise and fully connected encoder and decoder. It uses a multi-head attention algorithm to connect the encoder and decoder, and constructs representations of its input and output purely utilizing self-attention, without convolution or a sequence-aligned recurrent neural network~\cite{Transformer_2017}. Taking advantage of Transformer, BERT~\cite{bert_2018}, mainly pertained by masked language modelling (MLM), shows its great potential in dealing with NLP problems~\cite{sentence_bert_2019}. However, MLM effectively uses the bidirectional context of masked tokens, but it overlooks the dependence between the masked tokens~\cite{XLnet_2019}. Introduced by XLNet~\cite{XLnet_2019}, the permuted language modelling (PLM) is another powerful pre-training tool that captures the dependencies among tokens, but loses positional information of the full sentence~\cite{mpnet_2020}. To leverage the strengths of MLM and PLM and alleviate their shortcomings, masked and permuted language modelling (MPNet)~\cite{mpnet_2020} is proposed in 2020. In addition to considering the interdependence of the predicted tokens using PLM, MPNET also uses the positions of all tokens as input to enable the model to know the positional information for all tokens, which addresses the issues in BERT~\cite{bert_2018} and XLNet~\cite{XLnet_2019, mpnet_2020}. In addition, MPNet is pretrained on a huge text corpus, around 160GB of data.    

\begin{figure}
  \centering
  \includegraphics[width=.9\linewidth]{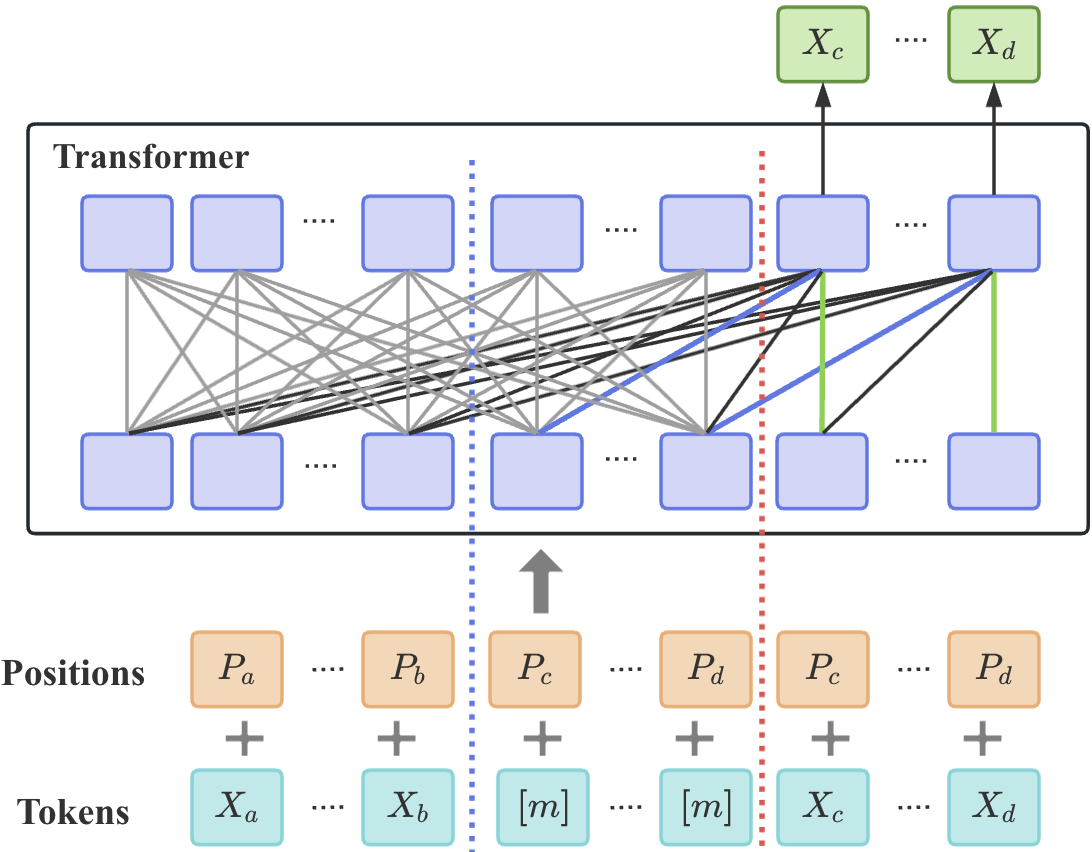}
  \caption{The Structural Overview of the MPNet Model. The leftmost inputs are non-predicted tokens and the rightmost tokens are selected as predicted tokens. The middle masks are the mask tokens of the predicted tokens. All the tokens along with their corresponding positions are fed into the model. Utilizing the MLM and PLM methods and the two-stream self-attention layers, the MPNet model extracts representations via a bidirectional model.}
  \label{fig:mpnet}
\end{figure}

To understand the mechanism of MPNet~\cite{mpnet_2020}, we first need to know how MLM and PLM work. While training, MLM masks predictable tokens and moves them to the rightmost position of the whole input sentence. For PLM, it will first permute tokens of the sentence and select the rightmost tokens as predicted tokens. Thus, both MLM and PLM have a unified view that the predicted tokens are all put at the end of input sentences~\cite{mpnet_2020}. Employing the techniques in MLM and PLM, MPNet~\cite{mpnet_2020} has the structure shown in Figure \ref{fig:mpnet}. MPNet tries to maximize the following objective while pre-training model $\theta$~\cite{mpnet_2020}. 
\begin{equation}
E_{z\in{Z_n}} = \sum_{t=c+1}^n \log{P(x_{z_t}\mid{x_{z<t}, M_{z>c}; \theta})}
\label{eqt:mpnet}
\end{equation}

As can be seen in Figure \ref{fig:mpnet}, $X_n$ is the input token and $P_m$ is the corresponding position information. The permuted input sequence is $X_a \ldots X_d$ and $c$ in Equation \ref{eqt:mpnet} denotes the number of non-predicted tokens. Thus, the predicted part is $X_c \ldots X_d$ and the remaining tokens $x_{z<t} = X_a \ldots X_b$ are the non-predicted. MPNet utilizes mask tokens [$m$] of the predicted part as input as well, corresponding to $M_{z>c}$ in Equation~\ref{eqt:mpnet}. Then, MPNet extracts representations from the input tokens ($X_a \ldots X_b$ [$m$] ... [$m$]), by bidirectional modelling~\cite{mpnet_2020, bert_2018}, shown by the grey lines in the Transformer in Figure \ref{fig:mpnet}. The blue and green lines in Figure \ref{fig:mpnet} represents the two-stream (content and query) self-attention~\cite{XLnet_2019} technologies that MPNet uses to predict tokens ($X_c \ldots X_d$)~\cite{mpnet_2020}. Leveraging the advantages of MLM and PLM, MPNet uses all the position information to obtain an overall view of the input~\cite{mpnet_2020}, which makes it an ideal language model to extract semantic/functional representations from instructions. 

We extract the first MPNet model from a pretrained sentence transformer published by Hugging Face~\cite{hugging_face_2016}, called \textit{all-mpnet-base-v2} model. \textit{all-mpnet-base-v2} is built based on the \textit{MPNet-base}~\cite{mpnet_2020} and fine-tuned on a one billion pairs dataset using a contrastive learning objective (the model is trained to identify which of a group of randomly picked other sentences was genuinely matched with a specific phrase in the dataset given a sentence from the pair)~\cite{all_mpnet_base_v2_2022}. It contains both \textit{MPNetTokenizer} and \textit{MPNetEmbedder}. 

\subsection{Subsequence Removal}

\begin{figure*}
  \centering
  \includegraphics[width=.9\linewidth]{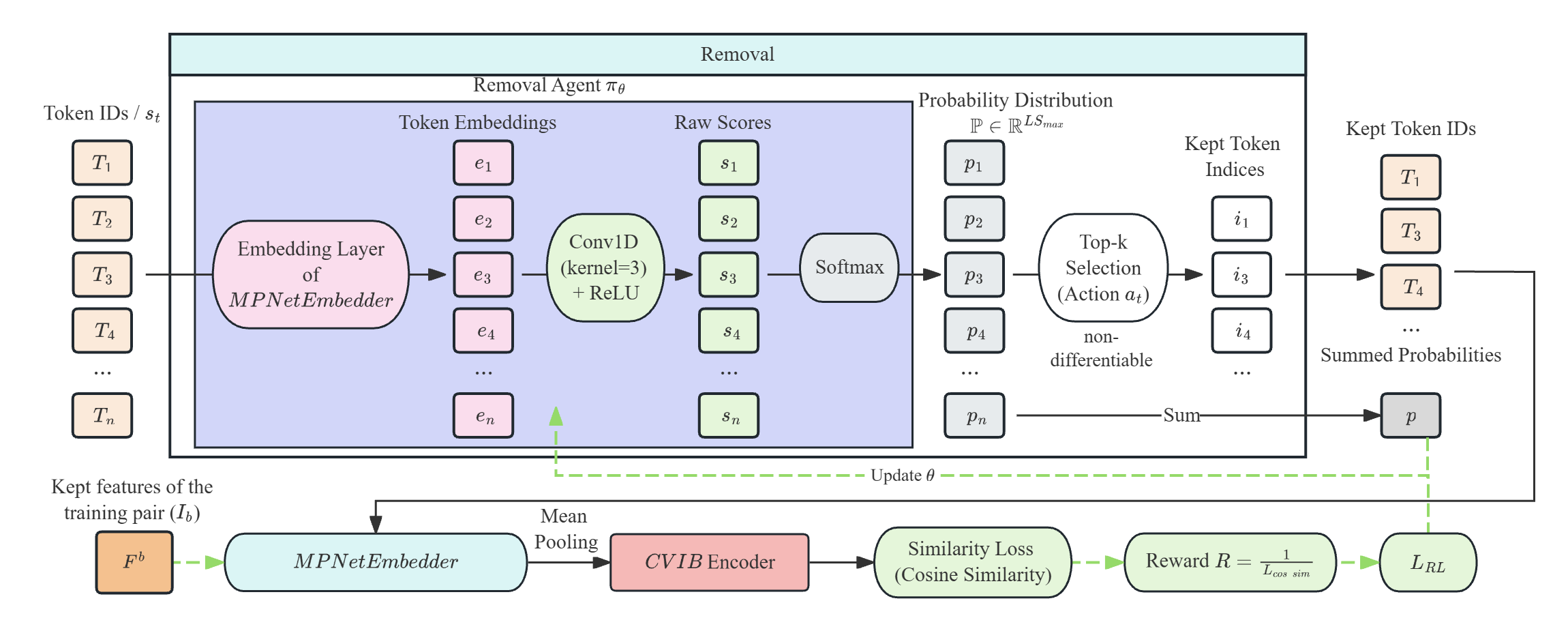}
  \caption{The Structural Overview of the \textit{Removal} Module. Input tokens are first fed into the \textit{Removal} agent $\pi_{\theta}$, which consists of an embedding layer, a 1D convolutional layer, and a Softmax function to generate token-level probabilities. The top-$k$ tokens with the highest probabilities are selected as the action, while their corresponding indices are used to obtain the retained tokens and their probabilities are summed as $p$. Green dashed paths denote training-only operations and solid black paths are shared by both training and inference..}
  \label{fig:removal}
\end{figure*}

Function inlining~\cite{agrawal_2020}, also known as function inline expansion or function inlining optimization, is a compiler optimization method that substitutes the actual body of the function for a function call at the call site during compilation~\cite{Chen_1993, karanpuria_2021}. This involves inserting the function's code at the call site rather than creating a separate function call instruction and jumping to the function's address in memory, therefore avoiding the cost of the function call and enhancing efficiency~\cite{karanpuria_2021, fish_2020, Irfan_2019}. Also, this method can remove the time and memory overheads that are caused when calling functions, due to the stacking of arguments, the saving and restoring of registers, and the jumping to and from the function~\cite{ibm_2018}. Moreover, some compilers only allow the inlining of functions that have already been defined in the file. Others first parse the full file so that functions defined after a call site may also be inlined. Yet other compilers permit the inlining of functions that have been declared in separate files~\cite{nullstone_2012}. Programmers have little or no control over which functions are inlined~\cite{Chen_1993}. Besides, compiler-injected code is another problem. Similar to functions inlining, compilers may replace or inject additional code into the original function during compilation. This is frequently done to incorporate specific language features or for optimization considerations. A compiler could inject extra code, for instance, to optimize memory utilization or execution time. Additionally, the compiler could add code to carry out bounds checking, memory access validation, or other security checks to protect the program.  

To remove as many noise tokens introduced by the above problems as possible, \textit{Removal}, a reinforcement learning (RL) based module, is placed between \textit{MPNetTokenizer} and \textit{MPNetEmbedder}. Reinforcement learning, as a branch of machine learning, is concerned with teaching an agent how to operate in a given environment in a way that would maximize a cumulative reward signal, known as a discounted reward, denoted by $G_t$. In reinforcement learning, an agent model performs an action in the environment and receives feedback in the form of rewards or punishments. The agent can learn an appropriate policy by the time the projected cumulative reward reaches the maximum over time~\cite{Kaelbling_1996}.

The agent is related to its environment in the traditional RL paradigm through observation/perception and action. Typically, the agent consists of a set of environment states, denoted by $S$, and a set of actions, denoted by $A$. Then, it receives an input, $o$, the observation of the environment's current state, $s$, at each stage of interaction~\cite{Kaelbling_1996}. The type of observation we employed is referred to as complete observation, which gives the agent comprehensive knowledge of the condition of the environment. The agent then decides which action ($a$) it produces as output and will get an immediate reward, $R$, once making an action. $R_a(s, s')$ defines the reward after transition from $s$ to $s'$ with action $a$. Moreover, the environment will typically alter as a result of the agent's action. The agent is informed of the immediate reward and the following $s'$ after making a decision, but is not informed of the action that would have been in its greatest long-term interest~\cite{Kaelbling_1996}. While making decisions about actions, the agent first distributes probabilities on each possible action and chooses the one with the largest probability. In this case, the agent actively gains relevant experience regarding the potential environmental states, actions, and rewards. Therefore, finding the policy $\pi$ for choosing actions that maximize a long-term reward signal is the aim of a reinforcement learning agent. Given the condition of the environment at time $t$, the policy is often expressed as a probability distribution across possible actions~\cite{Kaelbling_1996, Richard_1998}. The formula of the policy below

\begin{equation}
    \pi(a, s)=P(a_t=a \mid s_t=s)
\label{eqt:policy}
\end{equation}
shows the probability distribution $P$ if agent takes action $a$ given the states $s$ at time $t$.

For \textit{Removal}, it plays a \textit{one-step} ($t=1$) game, which means there are no iterations of actions, discounted factors, etc. Figure~\ref{fig:removal} illustrates the architecture of the \textit{Removal} module and its reinforcement learning process. The inputs, or observations/perceptions, of the agent model, are the environment states $o = s = S$. In this case, a tokenized instruction sequence, the token IDs $T={T_1,T_2,\ldots,T_n}$ serve as the state $s_t$ (the number of $T$ equals the max sequence length, denoted by $LS_{max}$, of \textit{MPNetTokenizer})The \textit{Removal} agent $\pi_\theta$, highlighted within the module, consists of the embedding layer of \textit{MPNetEmbedder}, a 1D convolutional layer (\textit{in\_channels=768, out\_channels=1, kernel\_size=3, padding="same"}) followed by ReLU, and a Softmax function. The embedding layer first maps each token ID $T_i$ to a 768-dimensional token embedding $e_i$. The convolutional layer then maps the token embeddings to raw scores $s_i$, which are normalized by Softmax to obtain a probability distribution $\mathbb{P}\in\mathbb{R}^{LS_{\max}}$. Each $p_i\in\mathbb{P}$ represents the estimated probability of retaining the corresponding token $T_i$. Based on $\mathbb{P}$, the \textit{Removal} module performs a one-step top-$k$ selection as the action $a_t$, where $k$ is the input sequence limit of the \textit{MPNetEmbedder}. The $k$ tokens with the highest probabilities are retained, and their corresponding indices are used to extract the kept token IDs and attention masks as the output features. Meanwhile, the probabilities associated with the selected tokens are summed to obtain $p$, which is used to construct the reinforcement learning objective. During training, as shown by the green dashed path in Figure \ref{fig:removal}, the kept features of the paired assembly sequence $I_b$, denoted by $F^b$, are processed together with the kept features of the current sequence by the shared \textit{MPNetEmbedder}. The resulting representations are mean-pooled and passed through the CVIB encoder, after which their cosine similarity loss $L_{\mathrm{cos_sim}}$ is calculated. The reward and the summed selection probabilities are subsequently used to calculate the reinforcement learning loss $L_{RL}$, whose policy gradient updates the parameters $\theta$ of the \textit{Removal} agent. Through this process, the agent learns to assign higher retention probabilities to tokens whose selection contributes to preserving the semantic relationship between paired assembly sequences, thereby filtering potentially noisy tokens.

\subsection{Cosine Similarity Loss}
To train \textit{Pluvio}, we adopt the cosine similarity loss~\cite{sentence_bert_2019} as one of our loss functions. 
In the NLP domain, it is frequently used to gauge textual similarity~\cite{han_2012}. The measure of cosine similarity is the cosine of the angle between the vectors, or the dot product of the vectors divided by the product of their lengths and the similarity degree value is always in the range [-1, 1]. The mathematical expression of cosine similarity given a pair of instruction embeddings, $E_a$ and $E_b$ that have $i$ components is~\cite{similarity_2017}
\begin{equation}
\begin{split}
    cos\_sim(E_a, E_b)&=\frac{E_a\cdot{E_b}}{\lVert{E_a}\rVert \cdot \lVert{E_b}\rVert}\\&=\frac{\sum_{i=1}^n{E_{ai}E_{bi}}}{\sqrt{\sum_{i=1}^n{E_{ai}^2}}\sqrt{\sum_{i=1}^n{E_{bi}^2}}}
    \label{eqt:cosine_sim}
\end{split}
\end{equation}

The loss function then computes the Mean Squared Error ($MSE$) to measure the difference between the actual label, $y_{actual}$, and $cos\_sim(E_a, E_b)$ as the cosine similarity loss~\cite{sentence_bert_2019}. The mean squared error is an assessor for classifiers or predictors, which evaluates the average squared variation between the vector value of the actual label~\cite{bickel_2015}. Our objective is to minimize the cosine similarity loss~\cite{sentence_bert_2019}, which we compute by
\begin{equation}
\label{eqt:cosine_sim_loss}
\begin{split}
    L_{cos\_sim}&=MSE(y_{actual}-cos\_sim(E_a, E_b))\\
    &=\lVert{y_{actual}-cos\_sim(E_a, E_b)}\rVert^2 \\
    &=\left\lVert {y_{actual}-\frac{E_a\cdot{E_b}}{\lVert{E_a}\rVert \cdot \lVert{E_b}\rVert}}\right\rVert^2
\end{split}
\end{equation}




\subsection{Reward Function and Policy Gradient}

As we discussed, in reinforcement learning, the agent model is informed of the immediate reward ($R$) after taking an action $a$~\cite{Kaelbling_1996}. Rewards are essential in reinforcement learning because they encourage agents to learn and explore. Without incentives, agents would have no means of recognizing which controls are crucial and they would be unlikely to progress in accomplishing their given job or resolving a particular control issue. We define the reward as 
\begin{equation}
R = \frac{1}{L_{cos\_sim}}
\end{equation}
However, in a one-step game, rather than the action iteration, the agent gives the first action as policy. Thus, the first immediate reward is the discounted reward ($G_t=R$).

The agent model (\textit{Removal}) cannot directly learn from the reward, because the reward value function is non-differentiable. Therefore, the loss function that is differentiable and related to the model parameters is essential to reinforce learning, since it can compute the policy gradient value for training. In general, the RL loss function is~\cite{torres_2021}
\begin{equation}
L_{RL} =-\sum_{t=0}^H{\nabla_\delta\log{\pi_\delta(a_t\mid{s_t})\cdot{G_t}}}
\end{equation}
which formulates the sum of the gradient of the logarithm of the policy (${\nabla_\delta\log{\pi_\delta(a_t\mid{s_t})}}$) with respect to the policy parameters $\delta$ and multiply with every step's discount reward ($G_t$) from the beginning until step $H$. Due to the input data being a pair of instruction sequences ($I_a$ and $I_b$), the output features of the \textit{MPNetTokenizer} with \textit{Removal} contain two token IDs ($T_a$ and $T_b$), attention masks ($M_a$ and $M_b$), and probability distributions ($p_a$ and $p_b$). Based on the input and Equation \ref{eqt:policy}, the RL loss function ($L_{RL}$) of a one-step game ($H=t=1$, $G_t=R$) can be computed by 
\begin{equation}
\begin{split}
L_{RL} &= -\beta_1\cdot {\nabla_\delta\log{\pi_\delta(a\mid{s})\cdot{R}}}\\
&=-\beta_1\cdot \log{\pi_\delta(a_{I_a}}\mid{s_{I_a}}) \cdot {\log{\pi_\delta(a_{I_b}}\mid{s_{I_b}})} \cdot {R} \\
&=-\beta_1\cdot \log{p_a} \cdot {\log{p_b}} \cdot {R}
\end{split}
\end{equation}
Moreover, in the RL loss, we add $\beta_1$ to control the learning rate of $L_{RL}$. After fine-tuning, the value of $\beta_1$ is finalized at 0.05.

While \textit{Removal} is designed to prune noise introduced by inlining and compiler-injected code, we acknowledge the risk of discarding critical semantic information. Two design choices mitigate this risk. First, \textit{Removal} is trained end-to-end with a reward ($R$), which is closely tied to the cosine-similarity objective (shown in the \textit{Equation 6}), so over-pruning of critical semantic tokens immediately reduces the reward and is discouraged by the policy gradient update (\textit{Equation 8}). Second, the agent (\textit{Removal}) performs prioritized retention by selecting the top-k tokens that have the $k$th largest probabilities ($p_a$ and $p_b$ in \textit{Equation 8}), rather than arbitrary truncation.

\subsection{Variational Information Bottleneck with Conditions}

As we discussed in Section~\ref{sec:related}, our goal is to find an appropriate way to build representations of assembly functions, which keep the maximal amount of useful information about functions and remove the nuisance information at the embedding level. We adopt the Information Bottleneck (IB)~\cite{Tishby_2000} theory in our work. Since we aim to implement a model that performs well for out-of-domain microarchitectures and libraries, we require the model to learn the key features of paired instruction sequences without the nuisance information from their microarchitectures, because this redundant information may negatively affect the performance of our model~\cite{Qian_2020}. Hence, using identical encoding as the stochastic representation of the input instruction sequences is not reasonable, even though this can solve the problems caused by the invariance of the mutual information to parameterizations~\cite{VIB_2017}. 
Instead, we use an information constraint, $C$, to constrain the mutual information $M_u$ between the encoding and the input source. We embed the tokens from \textit{Removal} to embeddings $E$ by the MPNet encoder as input sources and try to generate an encoding $Z$ that is maximally informative about learning target $Y$~\cite{VIB_2017}, so we have the objective function below.
\begin{equation}
    \max M_u(Z, Y)\:\: s.t.\: \: M_u(E, Z) \leq C
\end{equation}

Then, based on the Information Bottleneck method~\cite{Tishby_2000}, our goal is that the encoding $Z$ could be maximally informative representations of our learning targets $Y$ and maximally compress the redundant information in $E$~\cite{VIB_2017}. Therefore, a Lagrange parameter, $\beta_2$, is to control the trade-off: the larger the $\beta_2$ is, the more highly compressed representations will be generated~\cite{VIB_2017}. By this means, the IB objective function becomes
\begin{equation}
\label{eqt:IB}
    IB = M_u(Z, Y) - \beta_2 M_u(Z, E).
\end{equation}

In the above function, the $M_u(Z, Y)$ part simulates the $Z$ to be the representation with the most useful information of $Y$, and $\beta_2 M_u(Z, E)$ penalizes the representation for transmitting useless information from $E$~\cite{VIB_2017}. Alemi \textit{et al.}~\cite{VIB_2017} in 2017 proposed a practical way, called Variational Information Bottleneck (VIB), which seeks to acquire an effective latent representation of the input data for the downstream learning tasks by using variational inference~\cite{Qian_2020}. The VIB model proposed by Alemi \textit{et al.}~\cite{VIB_2017} has two essential components. One is a stochastic encoder, denoted by $p(Z|E)$ that converts input data into a low-dimensional representation and identifies a proper latent representation of the input data. The other one is a decoder, or a predictor, denoted by $q(Y|Z)$, that converts the low-dimensional representation back to the initial data space and is taught to recreate the input data from the representation while keeping the key features~\cite{Qian_2020, VIB_2017}. The major goal of variational approaches is to convert inference into an optimization problem. Based on the VIB method~\cite{VIB_2017}, the mutual information between encoding and the target $M_u(Z, Y)$ is defined as
\begin{equation}
    M_u(Z, Y) = \int dy \; dz \; p(y,z) \log \frac{p(y \mid z)}{p(y)} 
\end{equation}
where $z$ and $y$ are the instances of $Z$ and $Y$.

There is an intractable posterior probability distribution, $p(y|z)$ in our case. Variational approaches will attempt to find one variational approximation, $q(z|e)$, where $e$ is the instance of $E$. It is the tractable representation of the true posterior~\cite{kuleshov_2023}. Thus, known Kullback-Leibler ($D_{KL}$) divergence is always greater or equal to zero, the $D_{KL}[p(Y|{Z}),q(Y|{Z})]\geq 0$~\cite{VIB_2017} inequality equation can be written to
\begin{equation}
    \int dy \: p(y|z) \log p(y|z) \geq \int dy \: p(y|z) \log q(y|z)
\end{equation}

As it is an optimization problem, based on the above equations and empirical distribution function, the loss function for VIB~\cite{VIB_2017} is
\begin{equation}
\label{eqt:vib}
\begin{split}
    L_{VIB}=\frac{1}{N}\sum_{n=1}^N\left[\int dz\:p(z|e_n)\log q(y_n|z)-\right.\\
    \left.\beta_2 p(z|e_n)\log\frac{p(z|e_n)}{r(z)}\right]
\end{split}
\end{equation}
where $r(z)$ is the variational approximation to the marginal distribution of $Z$, and $N$ represents the number of inputs. We use a reparametrization technique~\cite{Kingma_2014} that samples from a noise distribution $\epsilon$ and then transforms it using a differentiable function dependent on the encoder network's parameters, resulting in a sample that approximates the desired distribution. The mathematical expression of the reparameterization trick on distribution $e$ by using the same deterministic neural network $f(e)$ used in the encoder is~\cite{Qian_2020}
\begin{equation}
    f(e)=\mu_z(e) + \sigma_z(e) \cdot \epsilon_z
\end{equation}
where $\mu_z$ and $\sigma_z$ are the mean and standard deviation of the distribution, respectively, and $\epsilon_z$ is a sample from a noise distribution~\cite{Kingma_2014}. 

This trick allows the gradients to be backpropagated through the encoder network, enabling it to be trained using stochastic gradient descent~\cite{VIB_2017}. Overall, the final objective function of VIB is~\cite{VIB_2017}
\begin{equation}
\label{eqt:objVIB}
\begin{split}
    J_{IB}=\frac{1}{N}\sum_{n=1}^N \mathbb{E}_{\epsilon\sim p(\epsilon)}[-\log q(y_n|f(e_n,\epsilon))] \\ +\beta_2 D_{KL}[p(Z|e_n),r(Z)]
\end{split}
\end{equation}

However, unsupervised learning is not well-suited for our task because while training, we lose control of VIB, so it may remove the useful information about instruction sequences and concentrate on the information about microarchitectures and optimizations. This situation varies depending on the data, which is contrary to our original goal (removing only information related to microarchitectures and optimizations from encoding). Therefore, in order to control the information which should be kept and which should be ``forgotten'', inspired by conditional Variational Auto-Encoder (CVAE)~\cite{CVAE_2018}, we add conditions, i.e., labels indicating the microarchitectures or optimizations of each paired assembly code to the VIB encoder. We thus named this new method Conditional Variational Information Bottleneck (CVIB). However, unlike CVAE, condition labels are only inputted during the training phase, and not used during testing. 

Similar to VIB, the encoder of the Variational Autoencoder (VAE)~\cite{Kingma_2014} receives source data and generates probability distributions in the latent space, while the decoder receives points in the latent space and returns artificial data. Consequently, for a given data sample, the encoder generates a distribution, a point in the latent space is sampled from this distribution, and this point is then transmitted to the decoder, which generates an artificial sample~\cite{dykeman_2016}. VAE decoder denoted by $Q(o|v)$ aims to reconstruct $o$ from the unobserved latent variables, $v$, and its encoder is thus denoted by $P(v|o)$. Hence, the flow in the VAE would be $o \rightarrow v \rightarrow o'$ where $o'$ represents the reconstructed data sample. Based on notions, the loss function of VAE can be defined as~\cite{CVAE_2018}
\begin{equation}
    L_{VAE} = \mathbb{E}_P[\log Q(o|v)] - D_{KL}(P(v|o)|Q(v))
\end{equation}

However, the decoder can only output artificial data samples that are randomized variants of the data samples upon which it was trained. Conditional VAE can give a specific artificial sample by introducing the condition labels, denoted by $l$, in both the encoder and decoder. So the system relies on both the latent space and the extra inputs $l$ to encode input data $P(v|l, o)$ and decode to the target value $Q(o|l, v)$. In this case, CVAE can generate artificial data samples according to the user's demands~\cite{dykeman_2016, CVAE_2018}. Mathematically, the variational objective of CVAE is formulated as~\cite{CVAE_2018}
\begin{equation}
\label{eqt:cvae}
    L_{CVAE} = \mathbb{E}_P[\log Q(o|l, v)] - D_{KL}(P(v|l, o)|Q(v|l))
\end{equation}

In our case, we do not use the VAE to generate fake assembly code embeddings. Instead, analogous to this method, components of the VIB model can be conditioned on a couple of observed labels (microarchitecture label $l_a$ and optimization label $l_o$) to learn/forget specific information from the given assembly embeddings. The difference between conditions in CVAE and CVIB is that conditions in CVIB are only used during the training phase, not the testing phase. In addition, we only add conditions in the \textit{CVIBEncoder} because the CVAE's goal is to reconstruct the input data sample, so it needs the conditions in the decoder to produce the full image by controlling the conditions. However, in our case, we focus on the representations $Z$ of the instruction embeddings $E$ in the VIB instead of the reconstructed $o'$ in the VAE. We need CVIB to purify the assembly code embeddings by forgetting the nuisance information of microarchitectures, libraries, and optimizations. Thus, we do not need the condition labels inputted to the \textit{CVIBDecoder}.

Hence, given the fact that our input is a pair of instruction embeddings encoded by \textit{MPNetEmbedder}, then $N=2$, based on Equation \ref{eqt:cvae}, \ref{eqt:objVIB} and \ref{eqt:vib} and the previous statement, we propose the CVIB (conditions in training only) information loss function as
\begin{equation}
\label{eqt:cvib}
\begin{split}
    L_{CVIB}=\frac{1}{2} \cdot \sum_{n=1}^2
    \left[ \int dz\:p(z|e_n, l_a, l_o)\log q(y_n|z) \right.\\
    \left. -\beta_2 \cdot p(z|e_n, l_a, l_o)\log\frac{p(z|e_n, l_a, l_o)}{r(z|l_a, l_o)}\vphantom{\int}\right]
\end{split}
\end{equation}
and the CVIB (conditions in training only) final objective function
\begin{equation}
\label{eqt:objCVIB}
\begin{split}
    J_{CVIB}=-\frac{1}{2}\cdot\sum_{n=1}^N \mathbb{E}_{\epsilon\sim p(\epsilon)}[\log q(y_n|f(e_n,\epsilon))]\\
    +\beta_2 \cdot D_{KL}[p(Z|e_n, l_a, l_o),r(Z|l_a, l_o)]
\end{split}
\end{equation}

\subsection{Semantic Search}
Semantic search is the query-answer searching technique that captures the meanings of query and entries to improve the performance of the model, as opposed to lexical search, which seeks exact matches of the query terms or similar phrases, without comprehending the context of the question~\cite{Hannah_2016}. Its goal is to find the most related answer to the query in a corpus of entries. For its mechanism, all of the entries in the corpus would be embedded into a vector space. Also, the query is embedded into the same vector space during the search, and the closet embeddings from the corpus to the query embedding are identified as the output answer because they have high semantic overlapping with each other~\cite{sentence_bert_2019}. Specifically, the semantic search utilizes the cosine similarity function to compute similarity scores between the query embedding and all embeddings of entries in the corpus. It then chooses the entry that has the highest similarity score as output~\cite{sentence_bert_2019}.

We adopt this method to measure the similarity score between two encodings of input assembly functions generated from \textit{CVIBEncoder}. We treat the first instruction embedding $Z_a$ as the query and the second $Z_b$ as the answer entry. Afterwards, instead of collecting the top entry, we treat its similarity score in the range [0, 1] as \textit{sscore}.

\section{Experiment}
\label{sec:exp}
\subsection{Data Preparation and Evaluation Setups}
To evaluate our models' performance under real-world circumstances, we collect open-source projects from different vendors/libraries, namely Busybox, OpenSSL, Putty, Curl, Coreutils, sqlite3, and ImageMagick, compiled for different microarchitectures (ARM, x86, MIPS, PowerPC, AMD64). We then use the IDA Pro disassembler~\cite{chris_2008} to generate the  \texttt{asm.json.gz} files which contain the assembly function code from the packages. Then, we build a File2Ins model to merge all instructions of an assembly function into a sequence and store the path, function name, and instruction sequence in a dictionary. Afterwards, we use different filters to categorize assembly functions in the same microarchitecture or from the same libraries into groups. Based on sorted assembly functions, we generate our pairwise training and out-of-domain tests. 

\begin{table}[]
\caption{The Distributions of Microarchitectures and Libraries in Training and Out-of-Domain Datasets}
\label{tab:datasets}
\begin{tabular}{l|l|l}
\hline
\multicolumn{1}{c|}{\textbf{Dataset}} & \multicolumn{1}{c|}{\textbf{architectures}}                & \multicolumn{1}{c}{\textbf{libraries}}                                    \\ \hline
Training                              & \begin{tabular}[c]{@{}l@{}}ARM, AMD64, \\ x86\end{tabular} & \begin{tabular}[c]{@{}l@{}}busybox, OpenSSL,\\ sqlite3\end{tabular}       \\ \hline
OOD-ARCH Testing                             & PowerPC, MIPS                                              & \begin{tabular}[c]{@{}l@{}}busybox, OpenSSL, \\ sqlite3\end{tabular}      \\ \hline
OOD-LIBS Testing                              & \begin{tabular}[c]{@{}l@{}}ARM, AMD64, \\ x86\end{tabular} & \begin{tabular}[c]{@{}l@{}}putty, coreutils, curl,\\  magick\end{tabular} \\ \hline
OOD-ARCH\&LIB Testing                        & PowerPC, MIPS                                              & \begin{tabular}[c]{@{}l@{}}putty, coreutils, curl,\\  magick\end{tabular} \\ \hline
\end{tabular}
\end{table}



\subsubsection{File2Ins}
In the assembly data extraction phase, we collect the sequence of instructions from each block and merge each instruction sequence into a long string following the address order. Also, when collecting assembly functions, we would skip the functions that have less than ten blocks because they are too small to provide sufficiently useful information to learn.

\subsubsection{Out-of-domain Tests}
In the last phase of the data preprocessing, we adopt the out-of-domain strategy and create three pairwise out-of-domain tests and eight sub-validations to evaluate our models. In statistics and machine learning, out-of-domain tests, often referred to as out-of-domain validation, are a method for evaluating how well a prediction model performs on data that was not utilized during its training phase~\cite{brownlee_2021}. Our goal is to assure models could have higher accuracy and Area Under the receiver operating characteristic Curve (AUC) values than other baselines. Given two assembly functions' information, we set the ground truth value either by 1 (if two functions are functional/semantic similar) or by 0 (they are functional/semantic dissimilar) based on the functions' names. Also, all datasets are balanced (the ratio of label 1 and label 0 is 1: 1). The details of microarchitectures and libraries used for training and out-of-domain testings are shown in Table \ref{tab:datasets}. When building our training dataset, we select assembly functions derived from libraries (BusyBox, OpenSSL, and sqlite3) and ARM, AMD64, and x86 as three assembly microarchitectures, using assembly functions derived from the same libraries and microarchitectures of PowerPC and MIPS for the out-of-domain microarchitecture test (OOD-ARCH), using assembly functions from different libraries (Putty, CoreUtils, Curl, and Magick) and from the same microarchitectures for then out-of-domain libraries test (OOD-LIB), and using assembly functions from different libraries (Putty, CoreUtils, etc.) and from different microarchitectures (MIPS and PowerPC) as out-of-domain microarchitecture and libraries test (OOD-ARCH\&LIB). 

Since we generate OOD-ARCH and OOD-LIB test sets by randomly selecting eligible assembly functions and pairing them together, paired data in these sets may have the same or different microarchitectures and optimization flags (O0, O1, O2, and O3). For instance, one pair of assembly functions in OOD-ARCH may both have x86 microarchitectures but different optimization flags, O1 and O0. Hence, we filter the OOD-ARCH and OOD-LIB datasets and split into four subsets by the same/different microarchitecture and same/different optimizations, marked as OOD-ARCH-sameA, OOD-ARCH-diffA, OOD-ARCH-sameO, OOD-ARCH-diffO, OOD-LIB-sameA, OOD-LIB-diffA, OOD-LIB-sameO, OOD-LIB-diffO.

\subsubsection{Obfuscation}

In real-world scenarios, developers obfuscate assembly code or binaries to conceal program intent and protect their intellectual property~\cite{Sebastian_16}. Obfuscation techniques, such as instruction substitution and data encoding, are widely adopted to increase the effort required by human analysts and clone detection tools~\cite{Cullen_03}. Therefore, to evaluate each module and \textit{Pluvio} itself under real-world cases, we employ an open-source obfuscator, named \textsc{ASM-encrypt}~\cite{ASMencrypt_2020}, which is designed for parsing assembly code and applying programmable rewrite rules. Its modular parser and rule-based interface enable fine-grained control over instruction-level transformations across different microarchitectures. We integrate this tool into our preprocessing pipeline to perform instruction substitution and data obfuscation on the OOD-ARCH\&LIB test dataset. \textsc{ASM-encrypt}’s~\cite{ASMencrypt_2020} rule engine allows matching and substituting assembly patterns using regular expressions and contextual templates, aiming to obfuscate assembly without changing semantics. Based on \textsc{ASM-encrypt}~\cite{ASMencrypt_2020}, we extend its functionality with new transformation rules for \textit{MIPS} and \textit{PowerPC} microarchitectures. Since these two microarchitectures differ in instruction semantics, register conventions (e.g., PowerPC conditional branches can test bits in the Condition Register, whereas MIPS integer branches such as \texttt{beq} compare general-purpose registers directly~\cite{IBM_PowerPC_Branch,MIPS32_ISA}), etc., we implement two obfuscators for each microarchitecture to remain semantics unchanged while obfuscating. 

At the instruction level, we substitute equivalent operations. Using one \textit{MIPS} assembly operand as an example, \texttt{move \$t0, \$t1} could be obfuscated to \texttt{addiu \$sp, \$sp, -4}; \texttt{sw \$t1, 0(\$sp)}; \texttt{lw \$t0, 0(\$sp)}; \texttt{addiu \$sp, \$sp, 4}, ensuring that operand semantics remain unchanged while the opcode distribution and n-gram patterns change. At the data level, we adopt arithmetic or bitwise encodings (e.g., replacing constants $v$ with $((v \oplus 9) \oplus 9)$) and convert ASCII literals to hexadecimal byte sequences. These obfuscation techniques collectively reduce feature alignment among semantically equivalent functions, making the text-based assembly clone search much harder to succeed. In our experiment, we obfuscate each assembly pair in the OOD-ARCH\&LIB set to obtain the OBFS-OOD-ARCH\&LIB.

\begin{table*}[]
\centering
\caption{Performance of Baseline Models and \textit{Pluvio} on Out-of-Domain Microarchitecture Tests with the Same and Different Microarchitectures (OOD-ARCH-sameA and OOD-ARCH-diffA)}
\label{tab:OOD_ARCH_same_diff_arch}
\begin{tabular}{l|ccccc|ccccc}
\hline
\multicolumn{1}{c|}{\multirow{2}{*}{\textbf{Model}}}                             & \multicolumn{5}{c|}{\textbf{\begin{tabular}[c]{@{}c@{}}out-of-domain microarchitectures \\ (same microarchitecture within a pair)\end{tabular}}} & \multicolumn{5}{c}{\textbf{\begin{tabular}[c]{@{}c@{}}out-of-domain microarchitectures \\ (different microarchitecture within a pair)\end{tabular}}} \\ \cline{2-11} 
\multicolumn{1}{c|}{}                                                            & \textbf{auc}              & \textbf{accu}              & \textbf{prc}             & \textbf{rcl}             & \textbf{f1}             & \textbf{auc}               & \textbf{accu}              & \textbf{prc}              & \textbf{rcl}              & \textbf{f1}              \\ \hline
\textsc{OM(ResNet7)}~\cite{Zeping_2020}  & 0.685                    & 0.634                     & 0.601                   & 0.609                   & 0.605                  & 0.505                     & 0.527                     & 0.524                    & 0.517                    & 0.520                   \\
\textsc{OM(BERT)}~\cite{Zeping_2020}  & 0.651                    & 0.581                     & 0.560                   & 0.554                   & 0.557                  & 0.479                     & 0.487                     & 0.503                    & 0.483                    & 0.493                   \\
\textsc{Order Matters}~\cite{Zeping_2020} & 0.706                    & 0.643                     & 0.627                   & 0.631                   & 0.629                  & 0.537                     & 0.510                     & 0.553                    & 0.532                    & 0.542                   \\
\textsc{ResNet11}~\cite{Zeping_2020}                                                                         & 0.531                    & 0.504                     & 0.499                   & 0.480                   & 0.489                  & 0.390                     & 0.401                     & 0.436                    & 0.432                    & 0.434                   \\
\textsc{PalmTree}~\cite{Li_2021}                                                                         & 0.688                    & 0.681                     & 0.844                   & 0.386                   & 0.530                  & 0.544                     & 0.543                     & 0.585                    & 0.556                    & 0.570                   \\
\textsc{SAFE}~\cite{Massarelli_2019}                                                                             & 0.500                    & 0.500                     & 0.000                   & 0.000                   & 0.000                  & 0.500                     & 0.500                     & 0.000                    & 0.000                    & 0.000                   \\
\textsc{Asm2Vec}~\cite{Steven_2019}                                                                          & 0.497                    & 0.553                     & 0.000                   & 0.000                   & 0.000                  & 0.505                     & 0.501                     & 0.545                    & 0.505                    & 0.524                   \\
\textit{MPNet}~\cite{sentence_bert_2019}                                                                & 0.844                    & 0.745                     & 0.761                   & 0.774                   & 0.787                  & 0.738                     & 0.665                     & 0.696                    & 0.664                    & 0.689                   \\
\textit{MPNet-QA}~\cite{sentence_bert_2019}                                                       & 0.853                    & 0.836                     & 0.805                   & 0.822                   & 0.813                  & 0.787                     & 0.717                     & 0.735                    & 0.625                    & 0.676                   \\
\textit{DistilRoBERTa}~\cite{sentence_bert_2019}                                                             & 0.859                    & 0.848                     & 0.840                   & 0.815                   & 0.827                  & 0.821                     & 0.739                     & 0.789                    & 0.711                    & 0.748                   \\
\textit{Distil-ML}~\cite{sentence_bert_2019}   & 0.851                    & 0.078                     & 0.755                   & 0.756                   & 0.755                  & 0.728                     & 0.669                     & 0.703                    & 0.678                    & 0.690                   \\
\textit{MiniLM}~\cite{sentence_bert_2019}                                                                & 0.804                    & 0.741                     & 0.733                   & 0.662                   & 0.696                  & 0.673                     & 0.633                     & 0.667                    & 0.649                    & 0.658                   \\ \hline
\textit{\textbf{Pluvio}}                                                                           & \textbf{0.902}                    & \textbf{0.866}                     & \textbf{0.844}                   & \textbf{0.860}                   & \textbf{0.852}                  & \textbf{0.876}                     & \textbf{0.795}                     & \textbf{0.803}                    & \textbf{0.827}                    & \textbf{0.815}        \\ \hline  

\end{tabular}
\end{table*}

\begin{table*}[]
\centering
\caption{Performance of Baseline Models and \textit{Pluvio} on Out-of-Domain Microarchitecture Tests with the Same and Different Optimizations (OOD-ARCH-sameO and OOD-ARCH-diffO)}
\label{tab:OOD_ARCH_same_diff_opt}
\begin{tabular}{l|ccccc|ccccc}
\hline
\multicolumn{1}{c|}{\multirow{2}{*}{\textbf{Model}}}                            & \multicolumn{5}{c|}{\textbf{\begin{tabular}[c]{@{}c@{}}out-of-domain microarchitectures \\ (same optimization within a pair)\end{tabular}}} & \multicolumn{5}{c}{\textbf{\begin{tabular}[c]{@{}c@{}}out-of-domain microarchitectures \\ (different optimization within a pair)\end{tabular}}} \\ \cline{2-11} 
\multicolumn{1}{c|}{}                                                           & \textbf{auc}              & \textbf{accu}              & \textbf{prc}             & \textbf{rcl}             & \textbf{f1}             & \textbf{auc}               & \textbf{accu}              & \textbf{prc}              & \textbf{rcl}              & \textbf{f1}              \\ \hline
\textsc{OM(ResNet7)}~\cite{Zeping_2020} & 0.693                    & 0.643                     & 0.627                   & 0.634                   & 0.630                  & 0.684                     & 0.636                     & 0.603                    & 0.612                    & 0.607                   \\
\textsc{OM(BERT)}~\cite{Zeping_2020} & 0.664                    & 0.603                     & 0.578                   & 0.569                   & 0.573                 & 0.653                     & 0.583                     & 0.5630                    & 0.557                    & 0.560                   \\
\textsc{Order Matters}~\cite{Zeping_2020}                                                                   & 0.715                    & 0.667                     & 0.648                   & 0.645                   & 0.646                  & 0.710                    & 0.649                     & 0.526                    & 0.636                    & 0.576                   \\
\textsc{ResNet11}~\cite{Zeping_2020}                                                                        & 0.540                    & 0.516                     & 0.510                   & 0.500                   & 0.505                  & 0.537                     & 0.509                     & 0.504                    & 0.486                    & 0.495                   \\
\textsc{PalmTree}~\cite{Li_2021}                                                                        & 0.748                    & 0.686                     & 0.924                   & 0.514                   & 0.661                  & 0.701                     & 0.653                     & 0.892                    & 0.384                    & 0.537                   \\
\textsc{SAFE}~\cite{Massarelli_2019}                                                                            & 0.500                    & 0.500                     & 0.000                   & 0.000                   & 0.000                  & 0.500                     & 0.000                     & 0.000                    & 0.000                    & 0.000                   \\
\textsc{Asm2Vec}~\cite{Steven_2019}                                                                        & 0.752                    & 0.505                     & 0.003                   & 0.000                   & 0.000                  & 0.496                     & 0.477                     & 0.000                    & 0.000                    & 0.000                   \\
\textit{MPNet}~\cite{sentence_bert_2019}                                                               & 0.844                    & 0.842                     & 0.832                   & 0.837                   & 0.821                  & 0.830                     & 0.765                     & 0.803                    & 0.729                    & 0.809                  \\
\textit{MPNet-QA}~\cite{sentence_bert_2019}                                                      & 0.803                    & 0.834                     & 0.806                   & 0.832                   & 0.819                  & 0.837                     & 0.862                     & 0.844                    & 0.836                    & 0.840                   \\
\textit{DistilRoBERTa}~\cite{sentence_bert_2019}                                                            & 0.872                    & 0.900                     & 0.822                   & 0.845                   & 0.833                  & 0.822                     & 0.814                     & 0.809                    & 0.869                    & 0.838                   \\
\textit{Distil-ML}~\cite{sentence_bert_2019}  & 0.831                    & 0.866                     & 0.844                   & 0.815                   & 0.829                  & 0.853                     & 0.782                     & 0.826                    & 0.739                    & 0.780                   \\
\textit{MiniLM}~\cite{sentence_bert_2019}                                                               & 0.837                    & 0.737                     & 0.825                   & 0.765                   & 0.794                  & 0.812                     & 0.735                     & 0.795                    & 0.667                    & 0.725                   \\ \hline
\textit{\textbf{Pluvio}}                                                                          & \textbf{0.922}                    & \textbf{0.913}                     & \textbf{0.905}                   & \textbf{0.914}                   & \textbf{0.909}                  & \textbf{0.895}                     & \textbf{0.870}                     & \textbf{0.888}                    & \textbf{0.861}                    & \textbf{0.874}        \\ \hline          
\end{tabular}
\end{table*}

\begin{table*}[]
\centering
\caption{Performance of Baseline Models and \textit{Pluvio} on Out-of-Domain Libraries Tests with the Same and Different Microarchitectures (OOD-LIBS-sameA and OOD-LIBS-diffA)}
\label{tab:OOD_LIBS_same_diff_arch}
\begin{tabular}{l|ccccc|ccccc}
\hline
\multicolumn{1}{c|}{\multirow{2}{*}{\textbf{Model}}}                             & \multicolumn{5}{c|}{\textbf{\begin{tabular}[c]{@{}c@{}}out-of-domain libraries \\ (same microarchitecture within a pair)\end{tabular}}} & \multicolumn{5}{c}{\textbf{\begin{tabular}[c]{@{}c@{}}out-of-domain libraries \\ (different microarchitecture within a pair)\end{tabular}}} \\ \cline{2-11} 
\multicolumn{1}{c|}{}                                                            & \textbf{auc}             & \textbf{accu}             & \textbf{prc}             & \textbf{rcl}            & \textbf{f1}            & \textbf{auc}              & \textbf{accu}              & \textbf{prc}             & \textbf{rcl}             & \textbf{f1}             \\ \hline
\textsc{OM(ResNet7)}~\cite{Zeping_2020}  & 0.693                   & 0.647                    & 0.617                   & 0.624                  & 0.620                 & 0.654                    & 0.632                     & 0.615                   & 0.560                   & 0.586                  \\
\textsc{OM(BERT)}~\cite{Zeping_2020}  & 0.661                   & 0.593                    & 0.578                   & 0.563                  & 0.570                 & 0.662                    & 0.591                     & 0.5764                   & 0.563                   & 0.570                  \\
\textsc{Order Matters}~\cite{Zeping_2020} & 0.715                   & 0.664                    & 0.639                   & 0.644                  & 0.641                 & 0.711                    & 0.667                     & 0.636                   & 0.599                   & 0.617                  \\
\textsc{ResNet11}~\cite{Zeping_2020}                                                                         & 0.546                   & 0.513                    & 0.517                   & 0.499                  & 0.508                 & 0.541                    & 0.506                     & 0.513                   & 0.467                   & 0.489                  \\
\textsc{PalmTree}~\cite{Li_2021}                                                                         & 0.510                   & 0.324                    & 0.456                   & 0.412                  & 0.433                 & 0.417                    & 0.319                     & 0.312                   & 0.480                   & 0.378                  \\
\textsc{SAFE}~\cite{Massarelli_2019}                                                                             & 0.500                   & 0.500                    & 0.000                   & 0.000                  & 0.000                 & 0.500                    & 0.500                     & 0.000                   & 0.000                   & 0.000                  \\
\textsc{Asm2Vec}~\cite{Steven_2019}                                                                          & 0.506                   & 0.506                    & 0.738                   & 0.506                  & 0.600                 & 0.411                    & 0.497                     & 0.000                   & 0.000                   & 0.000                  \\
\textit{MPNet}~\cite{sentence_bert_2019}                                                                & 0.838                   & 0.783                    & 0.836                   & 0.771                  & 0.802                 & 0.815                    & 0.796                     & 0.735                   & 0.766                   & 0.750                  \\
\textit{MPNet-QA}~\cite{sentence_bert_2019}                                                       & 0.880                   & 0.845                    & 0.841                   & 0.849                  & 0.845                 & 0.868                    & 0.834                     & 0.830                   & 0.847                   & 0.838                  \\
\textit{DistilRoBERTa}~\cite{sentence_bert_2019}                                                             & 0.816                   & 0.824                    & 0.848                   & 0.804                  & 0.825                 & 0.802                    & 0.819                     & 0.863                   & 0.782                   & 0.821                  \\
\textit{Distil-ML}~\cite{sentence_bert_2019}  & 0.854                   & 0.820                    & 0.830                   & 0.816                  & 0.823                & 0.871                    & 0.822                     & 0.838                   & 0.804                   & 0.821                  \\
\textit{MiniLM}~\cite{sentence_bert_2019}                                                               & 0.899                   & 0.081                    & 0.925                   & 0.817                  & 0.868                 & 0.857                    & 0.784                     & 0.740                   & 0.807                   & 0.772                  \\ \hline
\textit{\textbf{Pluvio}}                                                                           & \textbf{0.925}                   & \textbf{0.854}                    & \textbf{0.928}                   & \textbf{0.903}                  & \textbf{0.915}                 & \textbf{0.904}                    & \textbf{0.853}                     & \textbf{0.912}                   & \textbf{0.884}                   & \textbf{0.898}           \\ \hline      
\end{tabular}
\end{table*}

\begin{table*}[]
\centering
\caption{Performance of Baseline Models and \textit{Pluvio} on Out-of-Domain Libraries Tests with the Same and Different Optimizations (OOD-LIBS-sameO and OOD-LIBS-diffO)}
\label{tab:OOD_LIBS_same_diff_opt}
\begin{tabular}{l|ccccc|ccccc}
\hline
\multicolumn{1}{c|}{\multirow{2}{*}{\textbf{Model}}}                             & \multicolumn{5}{c|}{\textbf{\begin{tabular}[c]{@{}c@{}}out-of-domain libraries \\ (same optimization within a pair)\end{tabular}}} & \multicolumn{5}{c}{\textbf{\begin{tabular}[c]{@{}c@{}}out-of-domain libraries \\ (different optimization within a pair)\end{tabular}}} \\ \cline{2-11} 
\multicolumn{1}{c|}{}                                                            & \textbf{auc}             & \textbf{accu}             & \textbf{prc}             & \textbf{rcl}            & \textbf{f1}            & \textbf{auc}              & \textbf{accu}              & \textbf{prc}             & \textbf{rcl}             & \textbf{f1}             \\ \hline
\textsc{OM(ResNet7)}~\cite{Zeping_2020}  & 0.688                   & 0.640                    & 0.613                   & 0.622                  & 0.617                 & 0.664                    & 0.612                     & 0.573                   & 0.600                   & 0.586                  \\
\textsc{OM(BERT)}~\cite{Zeping_2020}  & 0.653                   & 0.596                    & 0.561                   & 0.555                  & 0.558                 & 0.638                    & 0.5662                     & 0.554                   & 0.527                   & 0.540                  \\
\textsc{Order Matters}~\cite{Zeping_2020} & 0.706                   & 0.656                    & 0.635                   & 0.633                  & 0.634                 & 0.694                   & 0.621                    & 0.508                   & 0.614                   & 0.556                  \\
\textsc{ResNet11}~\cite{Zeping_2020}                                                                         & 0.537                   & 0.501                    & 0.509                   & 0.492                  & 0.500                 & 0.511                    & 0.487                     & 0.489                   & 0.460                  & 0.474                  \\
\textsc{PalmTree}~\cite{Li_2021}                                                                         & 0.805                   & 0.785                    & 0.785                   & 0.666                  & 0.721                 & 0.578                    & 0.423                     & 0.929                   & 0.201                   & 0.330                  \\
\textsc{SAFE}~\cite{Massarelli_2019}                                                                             & 0.500                   & 0.500                    & 0.000                   & 0.000                  & 0.000                 & 0.500                    & 0.500                     & 0.000                   & 0.000                   & 0.000                  \\
\textsc{Asm2Vec}~\cite{Steven_2019}                                                                          & 0.503                   & 0.512                    & 0.787                   & 0.518                  & 0.625                 & 0.499                    & 0.293                     & 0.000                   & 0.000                   & 0.000                  \\
\textit{MPNet}~\cite{sentence_bert_2019}                                                                & 0.855                   & 0.812                    & 0.862                   & 0.824                  & 0.842                 & 0.822                    & 0.747                     & 0.835                   & 0.742                   & 0.786                  \\
\textit{MPNet-QA}~\cite{sentence_bert_2019}                                                       & 0.852                   & 0.836                    & 0.856                   & 0.862                  & 0.859                 & 0.803                    & 0.712                     & 0.821                   & 0.746                   & 0.782                  \\
\textit{DistilRoBERTa}~\cite{sentence_bert_2019}                                                             & 0.853                   & 0.871                    & 0.855                   & 0.795                  & 0.824                 & 0.899                    & 0.818                     & 0.861                   & 0.780                   & 0.819                  \\
\textit{Distil-ML}~\cite{sentence_bert_2019}  & 0.876                   & 0.821                    & 0.867                   & 0.813                  & 0.839                 & 0.898                    & 0.832                     & 0.852                   & 0.795                   & 0.823                  \\
\textit{MiniLM}~\cite{sentence_bert_2019}                                                                & 0.887                   & 0.870                    & 0.903                   & 0.849                  & 0.875                 & 0.866                    & 0.756                     & 0.851                   & 0.800                  & 0.825                  \\ \hline
\textit{\textbf{Pluvio}}                                                                           & \textbf{0.928}                   & \textbf{0.890}                    & \textbf{0.924}                   & \textbf{0.903}                  & \textbf{0.913}                 & \textbf{0.917}                    & \textbf{0.849}                     & \textbf{0.878}                   & \textbf{0.857}                   & \textbf{0.867}             \\ \hline    
\end{tabular}
\end{table*}

\begin{table}[]
\centering
\caption{Performance of Baseline Models \textit{Pluvio} on Out-of-Domain Microarchitectures and Libraries Test (OOD-ARCH\&LIB)}
\label{tab:OOD_ARCH_and_LIBS}
\begin{tabular}{l|lllll}
\hline
\multicolumn{1}{c|}{\multirow{2}{*}{\textbf{Model}}} & \multicolumn{5}{c}{\textbf{\begin{tabular}[c]{@{}c@{}}out-of-domain \\ microarchitecture and libraries\end{tabular}}} \\ \cline{2-6} 
\multicolumn{1}{c|}{}                                & \textbf{auc}         & \textbf{accu}         & \textbf{prc}         & \textbf{rcl}         & \textbf{f1}         \\ \hline
\textsc{OM(ResNet7)}~\cite{Zeping_2020}  & 0.497       & 0.528        & 0.523       & 0.390       & 0.447      \\
\textsc{OM(BERT)}~\cite{Zeping_2020}  & 0.485       & 0.516        & 0.522       & 0.358       & 0.425      \\
\textsc{Order Matters}~\cite{Zeping_2020} & 0.507       & 0.549        & 0.524       & 0.469       & 0.495      \\
\textsc{ResNet11}~\cite{Zeping_2020}   & 0.425       & 0.507        & 0.523       & 0.324       & 0.400      \\
\textsc{PalmTree}~\cite{Li_2021}                                                                         & 0.488       & 0.543        & 0.517      & 0.142       & 0.223      \\
\textsc{SAFE}~\cite{Massarelli_2019}                                                                             & 0.500       & 0.500        & 0.000       & 0.000       & 0.000      \\
\textsc{Asm2Vec}~\cite{Steven_2019}                                                                          & 0.499       & 0.500        & 0.000       & 0.000       & 0.000      \\
\textit{MPNet}~\cite{sentence_bert_2019}                                                                & 0.719       & 0.675        & 0.646       & 0.691       & 0.701      \\
\textit{MPNet-QA}~\cite{sentence_bert_2019}                                                       & 0.728       & 0.659        & 0.666       & 0.673       & 0.669      \\
\textit{DistilRoBERTa}~\cite{sentence_bert_2019}                                                             & 0.700       & 0.716        & 0.680       & 0.726       & 0.702      \\
\textit{Distil-ML}~\cite{sentence_bert_2019}  & 0.607       & 0.575        & 0.550       & 0.702       & 0.617      \\
\textit{MiniLM}~\cite{sentence_bert_2019}                                                                & 0.579       & 0.567        & 0.544       & 0.828       & 0.657      \\
\hline
\textit{CodeBERT}~\cite{feng_2020}                                                                & 0.526       & 0.571        & 0.663       & 0.204       & 0.312      \\
\textit{GPT-5.5}~\cite{openai_gpt55}                                                                &0.714 & 0.550 & 0.552 & 0.530 & 0.541      \\ \hline
\textit{\textbf{Pluvio}}                                                                           & \textbf{0.814}       & \textbf{0.760}        & \textbf{0.777}      & \textbf{0.789}       & \textbf{0.783}     \\ \hline

\end{tabular}
\end{table}

\subsection{Benchmark Setup}
Aiming to show the improvement \textit{Pluvio} achieved, we select several state-of-the-art approaches as baseline schemes, and we split some of them into sub-baselines by utilizing different combinations of components.

\textbf{Order Matters}~\cite{Zeping_2020}: This method pre-trains a BERT language model on the assembly code, passes the embeddings to a GRU-based Message Passing Neural Network (MPNN) layer, combines with the order embeddings from an 11-layer ResNet model on a CFG adjacency matrix, and get final embeddings from a fully connected layer for the final output.

\textbf{OM(BERT)}: This method is a sub-model of the \textsc{Order Matters} model~\cite{Zeping_2020}, which uses a basic BERT model without pre-training, passes the embeddings to a GRU-based Message Passing Neural Network (MPNN) layer, combines with the order embeddings from an 11-layer ResNet model on a CFG adjacency matrix, and get final embeddings from a fully connected layer for the final output.

\textbf{OM(ResNet7)}: This method is a sub-model of the \textsc{Order Matters} model~\cite{Zeping_2020}, which pre-trains a BERT language model on the assembly code, passes the embeddings to a GRU-based Message Passing Neural Network (MPNN) layer, combines with the order embeddings from a 7-layer ResNet model on a CFG adjacency matrix, and get final embeddings from a fully connected layer for the final output.

\textbf{ResNet11}: This method is a sub-model of the \textsc{Order Matters} model~\cite{Zeping_2020}, which uses only the order embeddings from an 11-layer ResNet model on a CFG adjacency matrix.

\textbf{PalmTree}~\cite{Li_2021}: This method adopts three pre-training tasks on the BERT model to capture the embeddings of assembly code to calculate the cosine similarity degree. 

\textbf{SAFE}~\cite{Massarelli_2019}: This method combines embeddings of assembly functions generated by the word2vec model and block embedding created by a self-attentive neural network to calculate the cosine similarity between assembly functions. 

\textbf{Asm2Vec}~\cite{Steven_2019}: This method generates embeddings of assembly code by using a PV-DM model.

\textbf{MPNet}~\cite{sentence_bert_2019}: The \textit{all-mpnet-base-v2} sentence transformer, built based on the \textit{mpnet-base} model and trained on a large and diverse dataset of over 1 billion training pairs, published by Hugging Face~\cite{hugging_face_2016}. We use this model to generate embeddings of the assembly code and compute function similarity degrees.

\textbf{MPNet-QA}~\cite{sentence_bert_2019}: The \textit{multi-qa-mpnet-base-dot-v1} sentence transformer model, built based on the \textit{mpnet-base} model and tuned for specific tasks, semantic search, by pre-training on a large and diverse set of question and answer pairs. We use this model to generate embeddings of assembly code.

\textbf{DistilRoBERTa}~\cite{sentence_bert_2019}: The \textit{all-distilroberta-v1} is one of the distilled versions of the RoBERTa-base models and trained on a large and diverse dataset of over 1 billion training pairs, totalling 82M parameters~\cite{distilroberta_base_hugging_face}. We use this model to generate embeddings of assembly code.

\textbf{Distil-ML}~\cite{sentence_bert_2019}: The \textit{distiluse-base-multilingual-cased-v2} is a Multi-Lingual model of Universal Sentence Encoder for 15 different languages (Chinese, Dutch, English, etc), which is published by Hugging Face~\cite{hugging_face_2016}. We use this model to generate embeddings of assembly code.

\textbf{MiniLM}~\cite{sentence_bert_2019}: The \textit{all-MiniLM-L12-v2} sentence transformer is trained on very large sentence-level datasets using a self-supervised contrastive learning objective, published by Hugging Face~\cite{hugging_face_2016}. We use this model to generate embeddings of assembly code and calculate function similarity.

\textbf{CodeBERT}~\cite{feng_2020}: \textit{CodeBERT} is a Transformer encoder (RoBERTa-style) pretrained on paired of natural language and source code corpora, with masked-language modeling and replaced-token detection, yielding contextual token embeddings amenable to mean-pooled sentence representations.

\textbf{GPT-5.5}~\cite{openai_gpt55}: Balancing model capability with inference cost and evaluation efficiency, we select \textit{GPT-5.5} as a general-purpose LLM baseline to assess whether its code-understanding capabilities transfer to assembly clone detection without task-specific fine-tuning. Given two assembly instruction sequences, we prompt the model to return a functional similarity score.

\subsection{Evaluation Metrics}
We choose to adopt the confusion matrix to evaluate the performance of our models, since it can precisely analyze the potential of a classifier by comparing the ground truth (0 and 1 in our task) and predicted labels (similarity degree score in the range of 0 and 1) of assembly code embeddings. The efficiency of the classifier is determined with regard to the following metrics generated from the confusion matrix entries~\cite{Pradeep_2021}. Accuracy is the percentage of correctly classified pairs; Precision is the proportion of correctly classified positive cases among all predicted positive cases; Recall is the proportion of correctly classified positive cases among all actual positive cases; then F1-score is a balanced measure (harmonic mean) of precision and recall. Overall, the higher the values, the better performance the model achieves.



Besides the four entries above, we also employ the area under a receiver operating characteristic (ROC) curve, abbreviated as AUC, as an important evaluator to measure the overall performance of our models~\cite{Melo2013}. Because its computation is based on the whole ROC curve and therefore takes into account all potential classification thresholds, the AUC is a reliable overall metric to assess the effectiveness of score classifiers. Typically, the AUC is determined by multiplying successive trapezoid regions beneath the ROC curve~\cite{Melo2013}. In contrast to accuracy, AUC ranks the predictions rather than just calculating their absolute values to quantify the quality of the model's predictions independent of the categorization threshold applied. Due to the classification-threshold-invariance and scale-invariance of AUC, it is a desirable evaluator to our task~\cite{Tom_2006, AUC_2022}.

\subsection{Overall Performance}
In the experiment, we evaluate baseline models and \textit{Pluvio}'s performance on eight different out-of-domain tests, and all testing sets contain balanced and pairwise assembly instruction sequences from different microarchitectures, libraries and optimizations. In Table~\ref{tab:OOD_ARCH_same_diff_arch}, the two columns of results, including Area Under ROC, accuracy, precision, recall and F-1 score, show the performance of the models on out-of-domain microarchitectures (OOD-ARCH), and their names are placed correspondingly on the left part. As we claimed, the filter is applied on the OOD-ARCH test to divide the test into two sub-tests, i.e., the same microarchitecture pairs and the different microarchitecture pairs. The middle column in Table~\ref{tab:OOD_ARCH_same_diff_arch} displays models' performances on the OOD-ARCH with the same microarchitecture pairs (OOD-ARCH-sameA) and the OOD-ARCH-diffA on the right column. Overall, all models perform worse in different microarchitecture pairs than in the same pairs, due to the inconsistency of microarchitecture types. Even so, \textit{Pluvio} still achieves the best performances (90.2\% and 87.6\% in AUC, 86.6\% and 79.5\% in accuracy) among all the models. The same happens in the remaining three OOD tests shown in Table~\ref{tab:OOD_ARCH_same_diff_opt}, Table~\ref{tab:OOD_LIBS_same_diff_arch}, and Table~\ref{tab:OOD_LIBS_same_diff_opt}.

\textit{Pluvio} achieves outstanding performance in every out-of-domain test, which demonstrates its effectiveness and robustness. By looking at our baseline and the state-of-the-art models, the series of Sentence Transformers~\cite {sentence_bert_2019} shows good performance, especially the \textit{all-mpnet-base-v2} model. This is because they consist of hundreds of MPNet layers and are pretrained on huge and diverse training pairs, over 1 billion diverse training datasets that equip Sentence Transformers~\cite{sentence_bert_2019} with high generalizability. However, they are designed as general-purpose models, not specifically for assembly code clone search. In contrast, \textit{Pluvio} is pretrained on only a trivial number of training data pairs compared to 1 billion and acquires greater generalizability (79.5\%, 87.0\%, 85.3\% and 84.9\% in accuracy in all four ``out-of-domain different microarchitectures and optimizations'' tests compared to 73.9\%, 81.4\%, 82.2\% and 83.2\%) than \textit{DistilRoBERTa}~\cite{sentence_bert_2019}. Then, for the \textsc{Order Matters}~\cite{Zeping_2020} model and its three sub-models, \textsc{OM(ResNet7)}~\cite{Zeping_2020}, \textsc{OM(BERT)}~\cite{Zeping_2020} and \textsc{ResNet11}~\cite{Zeping_2020}, achieve only at most 71.5\% in AUC among all eight sub-tests. In the training phase, \textsc{Order Matters}~\cite{Zeping_2020} extracts and combines the semantic-aware, structure-aware and order-aware embeddings from CFGs of assembly functions to measure the similarity degree, which improves the richness of embeddings information. This strategy disregards the intricate intra-instruction structures and relies heavily on the control flow, where contextual information is noisy and subject to influence from compiler optimizations~\cite{Li_2021}. Thus, plenty of noisy information from microarchitectures, libraries, and optimizations would mix into the embeddings, disturbing the model to lose focus. Out-of-domain tests clearly present this shortcoming in \textsc{Order Matters}~\cite{Zeping_2020}. Then, for the \textsc{PalmTree model}~\cite{Li_2021}, it realizes this problem, but only works on solving the impact of compiler optimizations and leaves the microarchitectures for their future work. Hence, from Table~\ref{tab:OOD_LIBS_same_diff_opt}, we could see that the \textsc{PalmTree model}~\cite{Li_2021} reaches 80.5\% AUC in the OOD-LIBS-sameO test, outperforming \textsc{Order Matters}~\cite{Zeping_2020} (70.6\% in AUC). However, in the out-of-domain tests with regard to microarchitectures, the \textsc{PalmTree}~\cite{Li_2021} model can only achieve 51.0\% and 41.7\% for the AUC in OOD-LIBS-sameA and OOD-LIBS-diffA tests. Focusing on obfuscations and not considering the impacts of different microarchitectures, libraries, and optimizations, \textsc{Asm2Vec}~\cite{Steven_2019} performs poorly in all OOD tests as well. For example, in the OOD-LIBS-diffA test in Table~\ref{tab:OOD_LIBS_same_diff_arch}, the Recall, Precision and F1-score of \textsc{Asm2Vec}~\cite{Steven_2019} are approaching zeros because nearly none of ``similar'' (label=1) class is classified correctly. \textsc{SAFE}~\cite{Massarelli_2019} has the same or even worse situation. It gives ``0s'' for every testing pair in all datasets. Since all testing sets are balanced (having 50\% similar (1) and 50\% dissimilar (0) testing pairs), \textsc{SAFE}~\cite{Massarelli_2019} achieves only 50.0\% in AUC and accuracy and 0.0\% in the remaining three metrics in all OOD tests. Due to being pretrained by over one million assembly code lines~\cite{Massarelli_2019}, \textsc{SAFE} is overfitting to a specific microarchitecture and loses its generalizability in all other microarchitectures, as well as for optimizations and libraries. Table \ref{tab:OOD_ARCH_and_LIBS} gives a more comprehensive view of all models' performance under the harder situation: out-of-domain microarchitectures and out-of-domain libraries. In this case, \textit{Pluvio} keeps the highest performance and robustness, achieving 81.4\% in AUC and 76.0\% in accuracy, and resulting in around a ten percent improvement compared to the existing state-of-the-art models. \textsc{CodeBERT}~\cite{feng_2020} is pretrained on source code with a subword tokenizer and masked language modeling that relies on identifiers and syntax absent in assembly. In this test, its embeddings lose transferable cues and collapse to a conservative operating point, leading to a very low recall (20.4\%) and AUC (52.6\%). The general-purpose LLM, GPT-5.5~\cite{openai_gpt55}, supports natural-language and code-related tasks. Their effectiveness for assembly clone detection across unseen microarchitectures and libraries requires task-specific evaluation, leading to relatively low AUROC values (71.4\%) and accuracy (55.0\%). As previously mentioned, \textit{Pluvio}’s design is specifically tailored for these challenges. \textit{Removal} and CVIB (conditions in training only) together remove noise from inlining functions and code injection at the tokenization and embedding levels. Due to the \textit{CVIB}, high domain adaptation makes key features maximally informative about given assembly functions. 

Furthermore, to evaluate the generalizability of our model and components, we conduct a complementary \textit{reverse-OOD-ARCH} test. We train \textit{Pluvio} exclusively on the OOD-ARCH test set (assembly pairs in microarchitectures of \textit{MIPS} and \textit{PowerPC}), and evaluate it on the training set (assembly pairs in microarchitectures of \textit{ARM}, \textit{AMD64} and \textit{x86}). In this setting, only the training pool and test pool are swapped, while every other setting remains identical.  \textit{Pluvio} achieves AUC of 89.1\% and accuracy of 87.5\% in \textit{reverse-OOD} test. Features learned from a narrower set of microarchitectures transfer to the original domain with only modest degradation, indicating that our model’s generalizability is not limited to a single pretraining dataset or specific microarchitectures.

Overall, \textit{Pluvio} consistently achieves outstanding performance in all tests, demonstrating its robustness and practicability in real-world assembly code clone search and all downstream works.

\begin{table}
\caption{Ablation Test for \textit{Pluvio} on Obfuscated Out-of-Domain Microarchitectures and Libraries. All models are trained by the balanced pairwise assembly functions training dataset and tested by the OBFS-OOD-ARCH\&LIB set.}
\label{tab:ablation}
\begin{tabular}{l|lllll}
\hline
\multicolumn{1}{c|}{\multirow{2}{*}{\textbf{Model}}}                            & \multicolumn{5}{c}{\begin{tabular}[c]{@{}c@{}}\textbf{out-of-domain} \\ \textbf{microarchitectures and libraries}\end{tabular}} \\ \cline{2-6} 
\multicolumn{1}{c|}{}                                                           & \textbf{auc}        & \textbf{accu}        & \textbf{prc}       & \textbf{rcl}       & \textbf{f1}       \\ \hline
\textit{MPNet}~\cite{sentence_bert_2019}                                                                & 0.534       & 0.557        & 0.480       & 0.501       & 0.490      \\
\begin{tabular}[c]{@{}l@{}}\textit{Pluvio -Removal -CVIB}\\ -Semantic search\end{tabular} & 0.640              & 0.578               & 0.635             & 0.512              & 0.568            \\
\textit{Pluvio -Removal -CVIB}                                                             & 0.667              & 0.582               & 0.660             & 0.534             & 0.590            \\
\textit{Pluvio-CVIB}                                                                     & 0.694              & 0.605               & 0.691             & 0.553             & 0.615            \\ 
\textit{\textbf{Pluvio}}                                                                           & \textbf{0.720}       & \textbf{0.662}        & \textbf{0.716}       & \textbf{0.537}       & \textbf{0.614}  \\ \hline   
\textit{Pluvio (10\% training)}                                                                     & 0.656              & 0.570               & 0.649             & 0.528             & 0.582            \\

\textit{Pluvio (+1 arch, +2 libs)}                                                                     & 0.763              & 0.704              & 0.753             & 0.602             & 0.670            \\\hline
     
\end{tabular}
\end{table}

\subsection{Ablation Test}
To assess the contribution of each component in \textit{Pluvio}, we conduct an ablation study on the most challenging evaluation set, the OBFS-OOD-ARCH\&LIB test, as shown in Table~\ref{tab:ablation}. In this table, the "-" symbol indicates the removal of the corresponding module. For instance, \textit{Pluvio-CVIB} denotes the model without the CVIB (conditions-in-training-only) module. The baseline MPNet~\cite{sentence_bert_2019} achieves 53.4\% AUC and 55.7\% accuracy under obfuscation, confirming that code obfuscation substantially reduces the separability learned by generic sentence encoders. In contrast, even the simplified \textit{Pluvio-Removal-CVIB-Semantic search} variant, without the Removal, CVIB, and semantic search modules, attains 64.0\% AUC and 57.8\% accuracy, demonstrating that the pairwise similarity learning framework captures more resilient structural cues from obfuscated assembly code. Reintroducing the semantic search extension and the Removal module further improves robustness. \textit{Pluvio-CVIB} achieves 69.4\% AUC and 60.5\% accuracy, proving the fact that \textit{Removal} successfully removes noisy tokens, such as redundant operations, introduced by code obfuscation and inlined or injected functions.  Finally, the complete \textit{Pluvio} achieves 72.0\% ACU with the support of the CVIB module, showing that CVIB in training successfully capture most informative representations of assembly code against obfuscation.

Moreover, to assess the sensitivity of \textit{Pluvio} to training data, we (i) decrease assembly function training pairs by using only one tenth of the original training pairs to train \textit{Pluvio}. In Table~\ref{tab:ablation}, the \textit{Pluvio (10\% training)} yields 65.6\% AUC and 58.2\% F1 (vs. 72.0\% and 61.4\% in the original model), a graceful degradation with the largest impact on recall. However, \textit{Pluvio} under high data scarcity still outperforms the first two ablations in the table, without collapsing or overfitting to noise. Then, we (ii) increase training diversity by including training pairs from one extra microarchitecture, \textit{MIPS}, and two more libraries, \textit{putty} and \textit{magick}, while keeping the total amount of training data the same as in the original set. As shown in Table~\ref{tab:ablation}, \textit{Pluvio (+1 arch, +2 libs)} improves performance to 76.3\% AUC and 67.0\% F1, with a notable precision gain (75.3\%). This is mainly because one of the microarchitectures and the two libraries present in the OOD-ARCH\&LIBS test were used in the training phase. On the other hand, broader training encourages the encoder to learn the key features of paired instruction sequences, without the nuisance information from microarchitectures/libraries. This is aligned with our model design that removes noise signals and enforces high domain adaptation, so that embeddings are maximally informative for similarity detection.

\subsection{Limitation}

\textit{Pluvio} shows performance degradation (AUC decreases from 90.2\% to 87.6\%) when comparing functions from OOD-ARCH, especially across different microarchitectures (OOD-ARCH-diffA test). In the OOD-ARCH\&LIB test, \textit{Pluvio}'s AUC further declines when doing clone search on both unseen microarchitectures and unseen libraries. Based on the observed results, we assume that \textit{Pluvio} might tend to degrade more severely on legacy architectures from older libraries, where instruction encoding, control-flow conventions, and memory-access models diverge from modern microarchitectures, and library-specific compilation patterns introduce distinct opcode and data-layout distributions that differ substantially from those seen during training. In addition, under code obfuscation (Table~\ref{tab:ablation}), \textit{Pluvio}’s AUC further drops to 72.0\%, suggesting that the encoder’s latent representation remains partially sensitive to instruction-level obfuscation, such as substitution and data encoding. These results indicate that the \textit{Removal} and CVIB modules still add robustness against noise introduced by inlined, injected, and obfuscated assembly functions. However, because these two modules operate at the token and embedding level, they have difficulty in removing syntax noise introduced by instruction substitution and data encoding obfuscations. Therefore, the model may lead to worse performance or even fail under extreme obfuscation.

\section{Related Work}
\label{sec:related}
Assembly code similarity detection has important cybersecurity applications. Traditionally, structural or syntactic features/representations such as CFGs from assembly code, and use them to match two assembly code fragments by calculating their similarity degrees. The shortcomings of these methods are their high computational cost and inefficiency. Recent techniques use an alternative way to extract representations from assembly functions. They automatically produce \textit{embeddings}, which are real-valued feature vectors~\cite{Irfan_2019}, from input assembly code. Xu \textit{et al.}~\cite{Xu_2017} propose a GNN-based model, called \textsc{Gemini}, which uses extracted graph embeddings for annotated CFG blocks (a CFG that has nodes annotated with particular fundamental block attributes~\cite{Irfan_2019}) of assembly functions for detection. However, the input features for this method are still manually selected, which is ineffective and time-consuming, especially for assembly code for different microarchitectures and from different vendors. Research attention is now shifting to semantic similarity, which captures if the functionalities of two compared assembly code fragments are identical~\cite{Irfan_2019}. Other techniques rely on instruction or raw byte co-occurrence rather than manually selected features~\cite{Irfan_2019}. 
 
Hence, people apply Natural Language Processing (NLP) to generate assembly code embeddings. In brief, word co-occurrence can be captured in word embeddings, so the embedding of a whole sentence can be built based on it. Zuo \textit{et al.}~\cite{zuo_2019} adopt an analogous strategy, considering a token as a word and a block as a sentence, and encode the semantic vector of a phrase using LSTM. Unlike the aforementioned approaches, Ding \textit{et al.}~\cite{Steven_2019} build \textsc{Asm2Vec}. It uses the PV-DM model~\cite{Le_2014} to concurrently produce instruction embeddings and an embedding for the function comprising these instructions to identify comparable functions in assembly code. \textsc{Asm2Vec}~\cite{Steven_2019} overcomes the problem in recent works that fail to consider the relationships between features and identify those unique patterns that can statistically distinguish assembly functions. However, hardware platforms and systems have various specifications and requirements. Thus, when assembly programs comply with different hardware platforms, they have different microarchitectures and are not compatible with each other. In this case, even though \textsc{Asm2Vec} leads a promising way to handle assembly code, it losses generalizability, since it fails to pass the cross-microarchitecture test. Massarelli \textit{et al.}~\cite{Massarelli_2019} utilize strategies in the NLP domain as well and propose the \textsc{SAFE} model, as an extended variation of \textsc{Gemini}. It uses the word2vec model to create the embedding of assembly functions and a self-attentive neural network to generate block embeddings. It was trained by millions of lines of assembly code. Although this method achieves higher efficiency and generalizability than prior work, as it detects stripped and cross-microarchitecture assembly code, it leaves much room to improve. Li \textit{et al.}~\cite{Li_2021} propose a BERT-based model called \textsc{PalmTree}, an assembly language model pretrained by three self-supervised tasks on large-scale datasets to generate embeddings of instructions for similarity detection. Unlike previous methods, it divides instructions into precise fundamental components. \textsc{PalmTree} outperforms previous methods in outlier detection and basic block similarity search, but it still fails to consider the real-world testing cases of assembly codes that are complied for different microarchitectures, vendors, and optimizations. Realizing this problem, Yu \textit{et al.}~\cite{Zeping_2020} propose a semantic-aware neural network, the \textsc{Order Matters} method. It is pretrained on three different levels of tasks. A BERT model extracts semantic-aware embeddings from blocks and is then followed by a GNN model that collects structural-aware embeddings. Moreover, they use a CNN model on adjacent matrices of CFGs to capture the order information as an order-aware embedding. At last, an MLP layer is used to concatenate all embeddings to compute the final graph embedding. The \textsc{Order Matters} method~\cite{Zeping_2020} outperforms many previous approaches and is somewhat robust for cross-microarchitecture. Nevertheless, it only supports two different platforms (x86 and ARM) and it does not fulfil our standards, since it still performs poorly in out-of-domain testing. \textsc{CodeBERT}~\cite{feng_2020} is pretrained on natural language and source code using masked language modeling and replaced-token detection. However, its pretraining signal and subword tokenizer are tuned to lexical and syntactic regularities, including identifiers and abstract syntax tree (AST) patterns, which are largely absent in assembly code. General-purpose LLMs such as \textsc{GPT-5.5}~\cite{openai_gpt55} support natural-language and code-related tasks. However, these capabilities do not necessarily imply robustness in transferring source-code language models to assembly analysis, including the low information density of assembly and substantial syntactic variation introduced by compiler optimizations~\cite{jiang2025nova}.

\section{Conclusion}
\label{sec:conclude}
In this paper, we propose \textit{Pluvio}, a robust and effective assembly code clone search engine. We combine the pretrained natural language model, transfer learning, reinforcement learning, and a novel method, which is a conditional variational information bottleneck, to create the embeddings from any given assembly function. \textit{Pluvio} introduces a new perspective to generate robust instruction embeddings by removing the nuisance information in the assembly functions. \textit{Removal} extracts a set of the top important tokens of the assembly function, so the impact of inlining functions and compiler-injected code can be mitigated. By conditional variational information bottleneck, \textit{Pluvio} is conditioned on learning the true semantics of assembly functions in the latent space and creates embeddings without nuisance information of microarchitectures, libraries, and optimizations. \textit{Pluvio} accepts not only the assembly functions, but also basic blocks, function segments, or even multiple lines of code as input without any extra information. We perform out-of-domain testing on assembly code clone search, utilizing a variety of microarchitectures, libraries, and optimizations. The convincing results demonstrate that \textit{Pluvio} is a robust and effective assembly code clone search engine which accurately classifies assembly codes in unseen microarchitectures or from unseen libraries.

\bibliographystyle{IEEEtran}
\bibliography{references}
\end{document}